\documentclass[manuscript]{aastex}

\usepackage{lscape}

\slugcomment{To appear in the Astronomical Journal}

\shorttitle{Asymmetric Thick Disk - Kinematics}
\shortauthors{Humphreys et al.}

\begin{document}

\title{Mapping the Asymmetric Thick Disk: III. The  Kinematics and Interaction with the Galactic Bar\altaffilmark{1}}

\author{Roberta M. Humphreys\altaffilmark{2,3},  
Timothy C. Beers\altaffilmark{4},
Juan E. Cabanela\altaffilmark{5,3}, 
Skyler Grammer\altaffilmark{2},
Kris Davidson\altaffilmark{2},
Young Sun Lee\altaffilmark{4},
Jeffrey A. Larsen\altaffilmark{6,3}
}

\altaffiltext{1}{Based on observations obtained at the MMT Observatory, a joint
facility of the Smithsonian Institution and the University of Arizona and at the Cerro Tololo Interamerican Observatory (NOAO)  operated by the
Association of Universities for Research in Astronomy (AURA).}
\altaffiltext{2}
{Astronomy Department, University of Minnesota, Minneapolis, MN 55455}
\altaffiltext{3}{Visiting Astronomer, Cerro Tololo Interamerican Observatory (CTIO), National Optical Astronomy Observatory (NOAO), which is operated by the 
Association of Universities for Research in Astronomy (AURA), Inc., under 
cooperative agreement with the National Science Foundation (NSF).}
\altaffiltext{4}
{Department of Physics and Astronomy and the Joint Institute for Nuclear Astrophysics,, Michigan State University, East Lansing, MI 48824}
\altaffiltext{5}
{Department of Physics and Astronomy, Minnesota State University Moorhead, Moorhead MN, 56563}
\altaffiltext{6}
{Physics Department, United States Naval Academy, Annapolis, MD 21402}

\email{roberta@umn.edu, beers@pa.msu.edu, cabanela@mnstate.edu, grammer@astro.umn.edu, kd@astro.umn.edu, lee@pa.msu.edu, larsen@usna.edu}

\begin{abstract}
In the first two papers of this series,  Larsen et al (2010a,b) describe our faint CCD survey 
in the inner Galaxy
and map the over-density of Thick Disk stars in Quadrant I (Q1) to 5 kpc or more along the 
line of sight. The regions showing the strongest excess are above  the density
contours of the bar in the Galactic disk.  In this third paper on the asymmetric Thick Disk,
we report on radial velocities and derived metallicity parameters for over 4000 stars in Q1, 
above and below the plane and in Q4 above the plane. We confirm the corresponding 
kinematic asymmetry first reported  by \citet{par04}, extended to greater distances 
and with more spatial coverage.  The Thick Disk stars in Q1 have a 
rotational lag of 60 -- 70 km s$^{-1}$ relative to circular rotation, and the Metal-Weak Thick Disk
stars have an even greater lag of 100 km s$^{-1}$. Both lag their corresponding 
populations in Q4 by $\approx$ 30 km s$^{-1}$. Interestingly, the Disk stars in Q1 also appear 
to participate in the rotational lag by about 30 km s$^{-1}$. The enhanced rotational lag for the 
Thick Disk in Q1 extends  to 4 kpc or more from the Sun. At 3 to 4 kpc, our sight lines extend above 
the density contours on the near side of the bar, and as our lines of sight pass directly 
over the bar the rotational lag appears to decrease.  This is consistent with a ``gravitational 
wake''induced by the rotating bar in the Disk which would trap and pile up stars behind it. We 
conclude
that a dynamical interaction with the stellar bar is the most probable explanation for the 
observed kinematic and spatial asymmetries.

\end{abstract}


\section{Introduction: The Asymmetric Thick Disk}

An excess of 
faint blue stars in Quadrant 1 (Q1) of the Galaxy compared to complemetary fields in Quadrant 4
(Q4) was initially recognized by \citet{lar96}. \citet{par03} subsequently extended the
survey to a much larger contiguous region covering several hundred square degrees. They confirmed
the original findings, mapped the stellar excess in Q1 from {\it l} $\sim 20 - 55\arcdeg$ and
{\it b} $\sim 20 - 45 \arcdeg$, and argued for a comparable asymmetry in Q1 below the plane.
The stars showing the excess were probable Thick Disk stars, 1 - 2 kpc from the Sun. 
\citet{par04} also reported an associated kinematic signature.  Velocities and 
metallicities of stars in 12 fields  in Q1 and Q4 showed  that the Thick Disk stars in
Q1 have a much slower effective rotation rate $\omega$, compared to the
corresponding Q4 stars, with  a significant lag of 80 to 90 km s$^{-1}$ in
the direction of Galactic rotation, greater than the expected lag of 30 to 50 km s$^{-1}$ 
\citep{Reid98,CandB,Carollo2010} for the canonical Thick Disk population.

Interpretation of the asymmetry, now referred to as  the Hercules Thick Disk Cloud \citep{lar08}, is
not straightforward. It is  tempting to assume that the asymmetry is the fossil remnant of a
merger, however the star counts were also consistent with a triaxial Thick Disk or Inner Halo,
with its axis in Q1, as well as with a gravitational interaction with the stellar bar in the disk \citep{1992ApJ...384...81W,2000MNRAS.317L..45H}. The latter is especially
intriguing given the corresponding asymmetry in the kinematics.  
A triaxial Thick Disk could also yield different 
effective rotation rates because of  noncircular streaming motions along 
its major axis. The star count excess appears to terminate near {\it l} $\sim$ 55$\arcdeg$ \citep{par03}. 
To search for evidence of triaxiality, in Paper I  \citep{lar2010a} we extended the star counts to fainter magnitudes, 
corresponding to greater distances, from {\it l} of 50$\arcdeg$ to 75$\arcdeg$.  
The fields at  55$\arcdeg$ to 75$\arcdeg$, show no significant 
excess, including the faintest magnitude interval, and therefore, do not support a 
triaxial interpretation of the asymmetry. 

The outstanding question relevant to the origin of the Hercules Cloud is its spatial extent
along the line of sight and its associated kinematics. In Papers I and II \citep{lar2010b} in this series we describe our faint CCD survey to map the spatial extent of the over-density. The photometric survey 
covers 47.5 square degrees in 63 fields in Q1 and Q4 above and 
below the Galactic plane. Except for fields with {\it b} $\sim$ 30 -- 40$\arcdeg$ in Q1, most of 
these regions are not covered by the Sloan Digital Sky Survey (SDSS, \citet{York}). 
\citet{lar2010b} find that the over-density  or star count asymmetry in Q1 extends to approximately 
 5 kpc along our line of
sight,  and that the regions showing the excess interestingly are above  the near side of  
the density contours for the bar in the Disk \citep{1992ApJ...384...81W}. We have also extended our corresponding spectroscopic survey to fainter magnitudes 
and greater spatial coverage. We have obtained additional medium-resolution spectra and 
now have  radial  
velocities and metallicity estimates for  more than 4000 stars in 31 fields. In the next 
section we describe the observations and data reduction. In \S {3} we discuss the kinematics
and confirm the asymmetry between the stars in Q1 and Q4, and in \S {4} we use the 
metallicity information from the spectra to separate the stars into the different 
populations. We determine the  rotational lag for the different populations in  
\S {5}, and in the concluding section we summarize the kinematic and spatial asymmetries in 
the Hercules Cloud and its interaction with the Galactic bar in the Disk.

\section{Observations and Data Reduction}

One of the goals of the spectroscopy program is to obtain more complete spatial coverage 
and extend the survey to greater distances than were available to \citet{par04}. In addition to the 
their original 12 fields, we added  13  fields in Q1 above and below the plane and 6 fields in Q4
above the plane. The distribution of the fields on the
sky is shown in Figure 1. The stars were randomly chosen from our photometric catalogs to have a wide
range in apparent magnitude. Most of the targets were selected to have extinction-corrected $B-V$ colors $\le 0.6$~mag. This color cut was
adopted to isolate a population dominated by Thick Disk and Halo stars. In the case of those
fields selected from the MAPS Catalog of the POSS I \citep{cab03}\footnote{http://aps.umn.edu}, we used an $O-E$ color of $\le$ 1.0~mag, corrected for interstellar extinction,  which
corresponds to $B-V$ $\approx 0.6$~mag.  

The spectra were obtained with the Hectospec Multi-Object Spectrograph (MOS) \citep{Fab98} on the 
MMT 6.5m telescope on Mt. Hopkins for the northern fields in 2007 and 2008 and with 
the Hydra MOS on the 
Blanco 4m at CTIO in 2006 and 2008 for the fields with declinations below -20$\arcdeg$.  
The stars selected for observation with the Hectospec range in V magnitude from 16 to 19~mag
and for Hydra from 15 to 18~mag; 1 to 2 magnitudes fainter than used in \citet{par04}; 
consequently, the total exposure times were long to get adequate S/N for the faintest stars, $\sim$ 10, for reliable velocities.  A summary of the 
observations is provided in Table 1.  

The Hectospec\footnote{http://www.cfa.harvard.edu/mmti/hectospec.html} has a 1$\arcdeg$ FOV and uses 300 fibers each with
a core diameter of 250$\mu$m subtending 1$\farcs$5 on the sky. We used the 600 l/mm grating 
with the 4800{\AA} tilt  yielding  $\approx$ 2500{\AA} coverage with 0.54{\AA}/pixel resolution
and R of $\sim$  1700. 
The Hydra MOS\footnote{http://www.ctio.noao.edu/spectrographs/hydra/hydra.html} has 138 2$\arcsec$ fibers with a 40$\arcmin$ FOV. We  used it with 
the SiTe 2K x 4K CCD detector
and the KPGL1 632 l/mm grating plus BG39 filter with the grating tilted for 4167{\AA} yielding 
0.58{\AA}/pixel resolution, R $\sim$ 1100,  and $\approx$ 2400{\AA} coverage. 
In 2006 we observed additional stars in several of the fields included in \citet{par04} to
increase the number of stars in some of the southern fields for comparison with their corresponding
fields in the north. These settings are thus the same that they used,  
but with a new  CCD detector.  
  
The Hectospec data was reduced using ESPECROAD\footnote{Developed by Cabanela in consultation with Doug Mink and other 
staff at CfA and available with documentation for download at http://iparrizar.mnstate.edu/juan/research/ESPECROAD}, a portable version of the CfA/SAO data reduction 
pipeline.  Hydra has its own data reduction package within the  Image Reduction 
and Analysis Facility software package (IRAF\footnote{IRAF is written and suppor
ted by the IRAF programming group at the National Optical Astronomy Observatories (NOAO) in 
Tucson, Arizona. NOAO is operated by the  Association of Universities for Research in 
Astronomy (AURA), Inc. under cooperative agreement with the National Science Foundation}). Both software packages  have  
multiple tasks for flat fielding, fiber throughput 
correction, wavelength calibration, spectrum extraction, and sky subtraction.
The separate exposures, often obtained on different nights, were coadded and cosmic
rays removed.
The radial velocities were measured with the IRAF task RVIDLINES.  We manually
identified 2 to 3 spectral lines and then relied on RVIDLINES to automatically
identify the remaining lines from a list we created including the  Balmer series, 
the Ca II  H and K lines, and strong metallic lines found in A-G-type dwarf
stars. We selected lines that could be clearly distinguished from the noise. 
Misidentified lines were rejected based on large velocity residuals. For the majority 
of the stars the velocity error is better than $\sigma_{helio}$ $\sim$ 5 km s$^
{-1}$.  Standard stars were also observed each night, and in all cases their measured 
velocities agree well with the published velocities.

Together with the 741 radial velocities from \citet{par04}, we now have 
velocities for 4151  stars, 1814 and 1210 in Q1,  above and below the plane, respectively,  
and  1127 in Q4. A supplemental catalog  
 is available on-line with all of the velocity data plus the metallicity
and atmospheric parameters described below. Tables A1 and A2 are described in the Appendix
with sample data. 

Figures 2 and 3 show histograms of the V magnitudes and B-V colors for the stars with
velocities in Q1 and Q4. The colors are corrected for interstellar extinction 
using the maps from \citet{Schlegel}. There  are fewer stars with velocities at 
the faintest magnitudes in Q4 because they were all obtained with Hydra. The color histograms demonstrate
that similar populations of stars were selected for velocities in the three regions.

We have determined metallicity estimates and  atmospheric parameters based on
procedures similar to those used  
for stars in the Sloan Digital Sky Survey (SDSS: York et al. 2000;
Abazajian et al. 2009). We applied  a newly developed version of the SEGUE Stellar
Parameter Pipeline (SSPP: Lee et al. 2008a,b; Allende Prieto et al.
2008), called the n-SSPP\footnote{The non-SEGUE Stellar Parameter Pipeline} suitable for application to medium-resolution spectra other than those taken by SDSS/SEGUE.  The n-SSPP uses
  the Johnson $V$ magnitude and
$B-V$ color, and/or a 2MASS (Cutri et al. 2003) $J$ magnitude and $J-K$
color, corrected for a interstellar absorption and
reddening, together with an estimate of the observed radial
velocity. The program  determines estimates of the primary
atmospheric parameters (T$_{eff}$, log g, [Fe/H], [$\alpha$/Fe]) and their
 errors, as well as estimates of distance, making use of a
subset of the procedures described in Lee et al. (2008a, 2010). 
 Note that it is not necessary that the input spectra be flux
calibrated, or continuum rectified. It is also not strictly necessary
to supply input colors, since the n-SSPP makes internal estimates that
can be used as needed, but due to possible degeneracies in the derived
parameters color information is certainly preferred.

We used the n-SSPP to obtain metallicity and atmospheric parameter estimates for 
all of our program stars with acceptable spectra, including the data from
\citet{par04}, which were reprocessed with the new pipeline so that all of the parameters
are on the same system. Rejected spectra include
those with too low S/N ratios, or other problematic behavior.
The input  $B-V$ colors are from our own CCD measurements, or 
estimated from the MAPS $O-E$ colors.  The $J$ magnitudes
$J-K$ colors are from the 2MASS Point Source Catalog, 
absorption corrected or de-reddened according to the Schlegel et al.
(1998) dust maps. The derived parameters include the metallicity measurements, [Fe/H] and [$\alpha$/Fe],
and the atmospheric parameters T$_{\rm eff}$ and log $g$. The errors in the derived parameters by the n-SSPP are very similar to
those that the SSPP claims. The typical errors of the SSPP are 141~K,
0.23 dex, and 0.23 dex for $T_{\rm eff}$, log $g$, and [Fe/H],
respectively, after combining small systematic offsets quadratically for
stars with 4500~K $\leq T_{\rm eff} \leq$ 7500~K (Lee et al. 2008a).
The distance from the Sun is derived
from the photometry and the atmospheric parameters and are accurate to 10\% to 20\% (Beers et al. 2000). The input quantities and derived parameters 
with their errors are
included  in Tables A1 and A2 described in the Appendix.
Figures 4 and 5 show the histograms for T$_{\rm eff}$ and log $g$ from the n-SSPP pipeline. There are 
approximately the same relative numbers of stars over the range of temperatures in each region. 
The few relatively warm stars, T$_{\rm eff}$ $>$ 7000K, are most likely horizontal branch stars
in the Halo. The similar log $g$ distributions also indicate that the greater majority of stars
in our sample are main sequence.

\section{The Kinematics}

Figures 6 and 7 show the normalized histograms of the measured velocities corrected to 
the Local Standard of Rest (V$_{LSR}$) and the angular velocity of rotation $\omega$ for 
Q1 above and below the plane and for Q4 above the plane. The heliocentric velocities were
corrected to the LSR using the solar motion ($u{\odot} = 9 km s^{-1}$, $v{\odot} = 12 km s^{-1}$, $w{\odot} = 7  km s^{-1}$) reported in  \citet{Ibata97} from {\it Hipparcos} data. 
$\omega$ was calculated using the standard equation with $R_{\odot} = 8$ kpc \citep{Reid98}, V$_{\odot} = 220$ km s$^{-1}$, and $\omega_{\odot} = 27.5$ km s$^{-1}$ kpc$^{-1}$.  
Both quadrants show a wide range of velocities with high velocity tails extending to negative
velocites in Q1 and to positive velocities in Q4. These high velocity tails are due to a 
non-rotating or slightly retrograde Halo relative to the LSR, and are expected to be 
negative in the direction of Galactic rotation in Q1 and positive in Q4. 
Due to the addition of fainter stars 
in our observations, we find a larger number of very high velocity 
Halo stars in both quadrants than \citet{par03}, including a few with V$_{LSR}$ $\ge \pm 
400$ km s$^{-1}$.  

The mean LSR velocities and $\omega$ for each direction are  in
Table 2,  excluding the obviously high velocity stars with  V$_{LSR}$ greater
than $\pm 200$ km s$^{-1}$. The velocity distributions in Q1 above (Q1A) 
and below (Q1B) the plane are very similar with essentially the same mean V$_{LSR}$ and $\omega$, 
while the kinematic asymmetry between Q1 and Q4 is quite apparent.
Assuming an axisymmetric Thick Disk that is rotating about the Galactic center, we would
expect to measure positive LSR velocities in Q1,  as these stars are moving away from us, and 
negative velocities in Q4 since the stars would be approaching. The velocities should be 
comparable for stars at similar distances and symmetric directions but of opposite sign.  
At typical distances from the Sun, of 2  to 3 kpc, corresponding to the mean V magnitudes, 
17 to 18  mag, and assuming that most of these stars are near the main sequence turnoff,
the expected  V$_{LSR}$ velocities in Q1, for example, should be 
$\approx$ 10--27 km s$^{-1}$.  
However, the Thick Disk is known to rotate slower or lag the Disk by up to 
50 km s$^{-1}$. The observed LSR velocities for the Thick Disk would be expected to show a
relative net shift with respect to the Disk velocities in both quadrants, to more 
negative velocities in Q1 and to more positive velocities in Q4. 
The stars in Q1  have a significant mean shift of 45 km s$^{-1}$ up to 
65 km s$^{-1}$ while  Q4 stars have a smaller shift and a mean V$_{LSR}$ more consistent 
with the expected velocities. 
The much greater shift in Q1  with respect to the expected  velocities shows a clear 
asymmetry between the two directions.  

The results for the angular velocity of rotation, $\omega$, betweeen 
the two quadrants illustrates the greater net lag with respect to Galactic rotation
in Q1. In these directions, in the inner Galaxy ({\it l} = 20--50$\arcdeg$) most of the 
stars will be at  distances from the Galactic center that range from 6 to 7.5 kpc. 
Using the power-law fit to the rotation curve in \citet{bb93}, 
stars at these Galactocentric distances would have an expected $\omega$ of  
30 to 39 km s$^{-1}$ kpc$^{-1}$. The mean $\omega$
for Q4, 31.3 $\pm$ 0.6 km s$^{-1}$ kpc$^{-1}$,  is thus marginally consistent with 
Galactic rotation with perhaps only a small lag. The stars in Q1, however, 
have a much slower effective 
rotation, with a mean $\omega$ of only 19.6 $\pm$ 0.5 km s$^{-1}$ kpc$^{-1}$,
thus confirming the kinematic asymmetry between Q1 and Q4. 

The velocities for the individual fields are in Appendix B. 
Four  of the fields in Q1 are paired directly  across the Galactic plane 
({\it b} = $0\arcdeg$),  and nine fields in Q4 have complementary fields in Q1 across
the {\it l} = $0\arcdeg$ line. In all four cases in Q1, the mean V$_{LSR}$ velocities and
rotation rates agree to well within the standard error while the matching fields
in Q1A and Q4 exhibit the slower rotation and greater lag in Q1.  
The three fields, H050+31, H050-31, and H310+31, illustrate the differences quite well.
H050+31 and H050-31 have remarkably similar mean V$_{LSR}$'s and $\omega$'s 
of $\sim$ -51 km s$^{-1}$ and 17 to 
18 km s$^{-1}$ kpc$^{-1}$, while the field at the same latitude and
complementary longitude,  H310+31, has a mean V$_{LSR}$ of +8.7 km s$^{-1}$ and 
mean $\omega$ of 26 km s$^{-1}$ kpc$^{-1}$. All three fields show slower effective 
rotation rates and a lag with respect to Galactic rotation, but it is much greater
in Q1.  
They  illustrate the kinematic symmetry in Q1 and asymmetry between Q1 
and Q4. 

The SDSS Data Release 7 overlaps our higher latitude fields in Q1 at {\it b} $\sim$ 30 --40$\arcdeg$.
As a check on our results, we queried the SDSS database from {\it b} $= 30 - 40\arcdeg$ 
and {\it l}  $= 20 - 50\arcdeg$. We transformed the SDSS magnitudes and colors to Johnson V and B-V
\citep{Jordi}, corrected them for interstellar extinction, and corrected the velocities to the LSR. 
Adopting our magnitude, color and V$_{LSR}$ ranges, the mean V$_{LSR}$ for 1747 stars in this
region is $-37.8 \pm 1.9$ km s$^{-1}$  with a mean $\omega$ of 16.9 $\pm$ 0.6 km s$^{-1}$ kpc$^{-1}$,
comparable to our results in Q1 for fields at similar latitudes. This result not only confirms
our results with independent data, but also demonstrates that the
rotational lag and presumed asymmetry extends across a large contiguous region in {\it l} and {\it b}.

One of the fields in Q1 below the plane, H048-45, has an anomalously 
negative velocity distribution and a large negative mean V$_{LSR}$ compared with the other fields 
in Q1. The other characteristics
of the stellar population, metallicities, colors, and magnitude distribution, appear to
be normal. The origin of its very negative velocity distribution is not known. 
It is not included in the subsequent analysis, but is discussed separately later.

The velocities and rotation rates discussed in this section 
represent a mixture of Disk, Thick Disk, and Halo stars\footnote{Some authors prefer old or 
thin disk. In this paper we simply use Disk to refer to this population. We also use Halo for 
what  is likely a
mix of Inner and Outer Halo stars in this paper.}. In the next section we use
the additional information on their metallicities to examine their kinematics as
a function of stellar population.

\section{Metallicities and Population Separation} 

Figure 8 shows the normalized metallicity distribution functions (MDFs) in the three regions for all of the stars with  [Fe/H] derived from the n-SSPP pipeline described in \S {1}.
The predominance of low-metallicity  stars in our 
sample confirms our color selection criteria. The three regions have very
similar MDFs with mean [Fe/H] of -1.14, -0.90, and -1.15 for 
Q1A, Q1B, and Q4, respectively. The mean error in [Fe/H] is 0.05 {\it dex} over all the
fields with a standard deviation of 0.04 {\it dex}. All three regions  have a
significant low metallicity tail.  Removing the high velocity stars with V$_{LSR}$ $\ge$ 
$\pm$ 200 km s$^{-1}$ does not significantly alter the mean [Fe/H] values.  Q1B  and Q4  
have several objects with positive values for [Fe/H]. For many of the fields in these two regions, the stars
for velocities were initially selected from the MAPS Catalog before the CCD photometry was available.
The more uncertain photographic colors may have led to more stars with redder B-V colors and  therefore
more Disk stars in the samples.  

In our previous study \citep{par04} we removed the high velocity stars and simply used [Fe/H] 
to separate the three primary populations, but recent studies of Disk and 
Thick Disk stars have demonstrated
that [Fe/H] alone is not sufficient to separate these two populations. 
Thick Disk stars primarily have [Fe/H] between $-0.5$ and $-1.2$
\citep{Wyse} although, a mix of
Disk and Thick Disk stars are now recognized with metallicities between 0 and $-0.5$
\citep{Bensby03,Nissen}.  
Furthermore, \citet{CandB} and \citet{Carollo2010} have argued for 
a Metal-Weak Thick Disk (MWTD) with somewhat lower metallicities, different kinematics, 
and a higher scale height than the canonical  Thick Disk. This additional population
 (-1.8  $<$ [Fe/H] $<$ -0.8) overlaps the expected metallicity distributions 
 of the   Thick Disk and the Inner Halo. The primary goal of this study is to investigate
 potential differences in the kinematics of the populations between Q1 and Q4. We must apply the same 
 selection criteria to all three regions. Consequently, we have chosen not to rely on their presumed 
 kinematic properties, which we already know are different,  to aid in the 
 population separation. Instead we use the abundance parameters, [Fe/H], [$\alpha$/Fe],
 in combination with  the stars' distances from the Galactic plane ($|Z|$), 
 to separate the four populations:
 Disk, Thick Disk, MWTD, and Halo.  Given the errors in the individual measurements for 
 these parameters from our moderate-resolution spectra, plus the natural spread and
 overlap among these populations, we do not expect a clean separation, especially between 
 Disk and Thick Disk, and Thick Disk and MWTD.  The high-velocity stars are excluded from the 
 population criteria  described below. 

{\it The Disk.} We initially assign to the Disk all  stars with [Fe/H] $\ge 0$  and within 
650 pc of the Galactic plane, twice the expected scale height of the old or thin Disk population, and assume that this group represents a relatively uncontaminated Disk population. 
However, some  Disk stars are also observed with [Fe/H] values as low as -0.5 to -0.7 and thus
overlap in metallicity with the Thick Disk population \citep{Bensby03,Bensby04,Nissen}.  
Due to their different enrichment histories from Type II and Type Ia supernovae, 
the ratio of the $\alpha$-elements 
to Fe, in combination with the [Fe/H] parameter, fortunately provides for a separation of 
Disk and Thick Disk stars with [Fe/H] less than solar (see review by \citet{FandB}). 
The plots of [$\alpha$/Fe] vs [Fe/H] for the three regions in Figure 9 
show the prominent bend or ``knee'' in the distribution at [Fe/H] 
$\approx$ $-$0.5 observed by other authors. The sloping distribution from [Fe/H] 
$\approx$ 0 to $-0.5$, however, is  a mix of  Disk and Thick Disk stars.  
Since Disk stars typically have [$\alpha$/Fe] values less than 0.2,  
and Thick Disk stars $\ge$ 0.3,  accurate abundances for the $\alpha$-elements, derived from 
high-resolution spectra, permit the separation of the Disk and Thick 
Disk in this region. The mean error in
[$\alpha$/Fe] in our data is 0.04 with a standard deviation of 0.012. 
With the uncertainties  
in the parameters from our more moderate-resolution spectra a clear separation is not feasible.   
Therefore, for the Disk stars in this metallicity range,
we  adopt [$\alpha$/Fe] $\le$ 0.2 with the additional requirement that $|Z|$ 
$<$ 650 pc.  Stars in the same parameter space with $|Z|$ $>$ 650 pc are assigned to the Thick Disk.  

{\it The Thick Disk.}  Stars with [$\alpha$/Fe] $>$ 0.2 and [Fe/H] from $-0.5$ to $-1.2$ are
assumed to belong primarily to the Thick Disk, but there will be some overlap with the MWTD at the 
low metallicity side of this range. Given our color cutoff and the effective temperature 
distribution, we expect that most of the Thick Disk stars are near the main sequence turnoff. 
Adopting the corresponding luminosity, M$_{v}$ $\approx +5.6$ mag, as a maximum luminosity, 
the faintest (m$_{v}$ $\sim$ 19 mag) and most distant Thick Disk 
stars would be expected to be no more than 5 kpc from the Sun. 
We therefore use this distance limit  to select a sample of probable Thick Disk stars.  
For those stars with [Fe/H] $<$ $-0.8$, we also add the requirement that $|Z|$ 
is less than 2.0 kpc, approximately twice the vertical scale for the nominal 
Thick Disk \citep{lar03},  to roughly separate them from possible MWTD stars which have a higher scale
height.  

These criteria for the Disk and Thick Disk leave two orphan populations: a.) stars with
[Fe/H] $>$ $-0.5$ and [$\alpha$/Fe] $>$ 0.2, and b.) stars with  [Fe/H] $-0.5$ to $-1.2$ 
and [$\alpha$/Fe] $<$ 0.2.
For  group a, stars with [$\alpha$/Fe] $>$ 0.3 are  assigned to the Thick Disk according to the
above criteria. 
For the few
with [$\alpha$/Fe] between $0.2$ and $0.3$, those with $|Z|$ 
$<$ 650 pc are placed with the Disk stars. Since there are very few Disk stars with 
[Fe/H] below -0.5, group b stars are treated  with the same criteria as the Thick Disk,  and   
 any in the lower metallicity range, [Fe/H] $<$ $-0.8$,  with $|Z|$ $>$ 2 kpc  are 
assigned to the MWTD.

For those stars that do not have an [$\alpha$/Fe] measurement, but with an [Fe/H] that would place them
in either the Disk or Thick Disk, we adopted the [Fe/H] separation from \citet{par04} at $-$0.3.
All with [Fe/H] between $-$0.3 and $-$0.5 and distances $\le$ 5 kpc were assigned to the
Thick Disk. Those  with [Fe/H] between 0 and $-$0.3 and $|Z|$ $<$ 650 pc are assumed to be in
the Disk, and those $|Z|$ $>$ 650 pc and closer than 5 kpc, in the Thick Disk.  

{\it The  Metal-Weak Thick Disk.} \citet{Carollo2010} conclude that the MWTD contributes a
significant number of stars in the metallicity range  $-1.8 <$ [Fe/H] $< -0.8$ which
also  includes Thick Disk and Inner Halo stars. They also find that
the MWTD has a much higher scale height than the Thick Disk. 
Therefore for stars with [Fe/H]
from $-0.8$ to $-1.2$, we adopt a  $|Z|$ distance of 2 kpc for an approximate separation 
of the MWTD and the Thick Disk as described above. 
Stars with [Fe/H] $<$ $-1.2$ and $|Z|$ $>$ 4 kpc are assigned to the Halo. For this study we do not attempt to separate the Inner and Outer Halo. All stars
with [Fe/H] $<$ $-1.8$ are assigned to our Halo population. 

The population criteria are summarized in Table 3. Given our adopted criteria together
with the resolution of the abundance parameters, there
will undoubtedly be some contamination and overlap among these groups. 
For this reason we use V$_{LSR}$ to further refine the  population separation. There are  
several stars in the Thick Disk and MWTD samples with V$_{LSR}$'s close to our high-velocity cutoff. Most 
of these same stars have [Fe/H] values that would place them in the population overlap. For that reason we
assigned stars with V$_{LSR}$ $\ge$ 150 km s$^{-1}$ to the Halo. The resulting  mean values with their errors 
plus the mean [Fe/H] for each 
population are given in Table 4. The high-velocity stars
with V$_{LSR}$ $>$ 200 km s$^{-1}$ are also listed in Table 4. They are presumably 
members of the Halo but were not included in the four population groups.

\section{The Kinematic Asymmetry and Rotational Lag}

The  mean velocities and rotation rates between Q1 and Q4 in Table 4 for  
the Thick Disk, MWTD, and  Disk stars confirm the symmetry in the kinematics in Q1 above and below
the plane, as well as  
our previous  conclusion  that a significant population of Q1 stars are 
rotating slower than those in Q4.  
In Q1 the shift to more negative velocities and the rotational lag is quite apparent
for the Thick Disk population, confirming the kinematic
asymmetry between Q1 and Q4.  The Disk population in Q1 
also appears to be participating in the rotational lag to some extent.
The Q1 Disk stars above the plane have a significant negative $V_{LSR}$ compared to the 
slightly positive velocity for the stars below the plane.  
However,  as we noted previously,  there are significantly more Disk stars 
in Q1B  and  in Q4 which may yield a more representative result for these two regions.
Nevertheless, both regions in Q1 exhibit a  slower rotational rate ($\omega$)  relative  
to Q4.   
Given our population selection criteria, however, there is the possibility that a number of 
Thick Disk stars are included within this group.  As a test, we restricted our Disk population to 
stars with [Fe/H] $> 0$ and  within 2 kpc of
the Sun to get a sample more likely to be Disk stars only.  We obtain a negative mean $V_{LSR}$
of -8.9 $\pm$ 7.5 km s$^{-1}$ in Q1. Although there were only 21 stars and the uncertainty is large,
this result supports our conclusion that the Disk stars in Q1 also show the kinematic 
asymmetry. 

The MWTD shows the same asymmetry between 
Q1 and Q4  and evidence for an even larger lag with a much smaller rotational 
rate than for the Thick Disk. The results for the MWTD in Q4  also suggest a greater
lag, or shift to more positive velocities and smaller $\omega$, relative to the 
corresponding Q4 Thick Disk stars.  The MWTD stars also have a  higher 
velocity dispersion than the Thick Disk.  The results for the Halo stars are much as we would expect for 
a slowly moving,  retrograde population with a high velocity dispersion with respect to the LSR. 

The results for the individual fields are in Appendix C. The four paired fields across the 
{\it b} $= 0$ line yield self-consistent results for the populations, demonstrating the
 symmetry in Q1.
Earlier we mentioned what appeared to be 
an anomalously negative velocity distribution for H048-45 which one can see reflected in the 
numbers for this field   in Appendix C. The other characteristics of this field, such as its 
metallicity distribution, appear to be normal. The region of the sky containing this field is included in the SDSS-DR7
and SEGUE1. We examined the data in the direction of H048-45 and found several stars, selected by our criteria,
with the same high velocities.  On this basis, we conclude that our measured velocities are not in 
error or due to instrument or calibration errors. Furthermore,  our line of
sight toward  H048-45 does not appear to intersect a known cluster, dwarf galaxy or star stream, 
but given the abundance of fossil remnants and streams in the Halo, this field is not included in the above
analysis or the remaining discussion. Its inclusion would not alter the conclusions. 

 To determine the actual rotational lag, we must measure the velocity component in the 
direction of Galactic rotation, or the circular velocity. 
The observed V$_{LSR}$ is a combination of the stars' motions in the radial (v$_{r}$), tangential
 (v$_{\phi}$), and vertical (v$_{z}$) directions. Assuming that the stellar populations 
  have a common or shared motion in their respective regions, we use 
 the Levenberg-Marquardt technique \citep{Lev,Marq} for non-linear least squares with the 
 equation 

 \begin{equation}
 V_{LSR} = [v_{r}cos\theta + R_{\ast}(\frac{v_{\phi}}{R_{\ast}} - \frac{V_{\odot}}{R_{\odot}})sin\theta]cosb + v_{z}sinb
 \end{equation}

 where  $\theta$ is the angle between the line of sight from the Sun to the star 
 and the Galactic Center to the star, $\sin\theta = R_{\odot}{\sin}l/R_{\ast}$, 
 and R$_{\ast}$ is the distance of the star from the Galactic Center. 
 We then solve for v$_{\phi}$, v$_{r}$, and v$_{z}$ for the stellar populations in Q1 and Q4. 
 The results are summarized in Table 5. 
The solutions with all three parameters often gave results,  especially for v$_{z}$, that were 
not realistic, ranging from -50 to +84 km s$^{-1}$, and with large errors. We concluded that 
v$_{z}$ is not sufficiently constrained by our data and repeated the solutions with two 
parameters setting $v_{z} = 0$. We also include the results solving only for v$_{\phi}$. 
A few of the two-parameter solutions yielded what we presume are anomously large values
for v$_{r}$, up to 40  km s$^{-1}$  with large errors, although, in most cases, 
 v$_{\phi}$ was consistent with the three- and one-parameter solutions. 
The results for the Disk in Q1B did not converge. Although there are more Disk
stars than in Q1A, they are all in three fields at {\it b} $= 20\arcdeg$, and there was 
insufficient leverage in latitude for a  solution. The spatial coverage in Q1B is less 
uniform than in the other two regions, and the results are weighted by the three fields near 
the plane.  We therefore include a combined result for Q1 in Table 5, and in the following 
discussion we emphasize the  comparison of  Q1 and Q1A with Q4.  

The results for v$_{\phi}$, for the Thick Disk and MWTD in both quadrants, exhibit the rotational lag
with respect to circular rotation in the Disk, V$_{\odot}$ $\sim$ 220 km s$^{-1}$, and relative 
to their respective solutions for the Disk.  The solutions for the Thick Disk in Q1 are robust,
irrespective of the 
number of parameters, and indicate a rotational lag of 60 to 70 km s$^{-1}$ relative to circular
rotation and 20 to 30 km s$^{-1}$ slower than the Thick Disk stars  in Q4, which have a net lag 
of $\approx$ 40 km s$^{-1}$, as expected for the canonical Thick Disk. The MWTD stars in Q1 have a  
 significantly greater rotational lag of $\approx$ 100 km s$^{-1}$, 30 km s$^{-1}$ slower than 
 for the Thick Disk. The  solutions in Q4 for the MWTD, however,  suggest a smaller lag 
 of $\approx$ 60 km s$^{-1}$, again about 30 km s$^{-1}$ slower than the  corresponding 
 Thick Disk population, and 40  km s$^{-1}$ less than for the MWTD in  Q1. 
 The results also confirm our earlier suspicion that the Disk stars in Q1  participate
 in the rotational lag in the first quadrant. They appear to lag circular rotation by $\approx$ 
 30 km s$^{-1}$ while the Q4  stars have the expected circular rotation for the Disk. 

{\it  In summary, although there is considerable range in some of the solutions, the results 
 confirm the kinematic asymmetry and slower rotational lag in Q1.  All 
 three populations  lag their corresponding populations in Q4 by $\approx$ 
 30 km s$^{-1}$.}

In the next section, we conclude with a comparison of  the kinematic and spatial asymmetries in the  
 Hercules Thick Disk Cloud and its relationship to the bar in the Disk. 

\section{The Hercules Cloud and the Kinematic Asymmetry: Interaction with the Bar} 

In Paper II, \citet{lar2010b} confirm the over-density in Q1 among the faint blue stars
and argue convincingly that it is due primarily to the Thick Disk stars. The star count excess 
is strongest above the plane between longitudes 20$\arcdeg$ to 50$\arcdeg$ and at higher latitudes.
They conclude that, while still present at the lower latitudes, the signature is weaker, most likely 
due to the increased contribution from the Disk  along the line of sight. 
One of the goals of the deeper photometric survey was to map the asymmetry along the line of sight to determine
its full spatial extent. \citet{lar2010b} find that they can indeed identify the feature's far side
at about 5 kpc from the Sun, where the star count ratios between Q1 and Q4 return to near one along
the same sight lines. Significantly, the regions showing the strongest  excess   
are  over the near side of the density contours of the bar in the Disk  and appear to be associated with the 
increasing density of the bar along the sight lines.

The stars included in our velocity survey were selected by the same criteria in color and magnitude
range, and in most cases from the same fields in {\it l} and {\it b} on the sky as the 
stellar population
exhibiting the over-density. These are the same populations.
Thus, the stars showing the excess also participate in the kinematic asymmetry and 
the rotational lag.  Comparing the kinematic results for the individual fields in Appendix C, we 
find that the asymmetry and 
lag for the Thick Disk stars exists across the entire region, in Q1 above and below the plane. Paired 
fields across the {\it b} $= 0$ line give very similar results. The kinematics thus parallel 
the star count excess. Compared with the over-density in Paper II, 
the kinematic asymmetry extends
slightly beyond its apparent boundaries to 60$\arcdeg$ in two fields both at 20$\arcdeg$ in latitude.

The star counts suggest that we  see {\it through} the Hercules Cloud to its far side at distances 
on the order of 5 kpc or greater. Although, our observations of the Thick Disk stars 
are limited to within 5 kpc of the Sun, and   
there are few stars at distances beyond 4 kpc, we also examined the apparent lag as function of
distance along the line of sight. Figure 10 shows the variation in the mean V$_{LSR}$ with distance.  These two 
quantities appear to be correlated for large $r$. The mean V$_{LSR}$ appears 
to shift to  less negative values at greater distances, but the samples are too 
small to quantify the relation well.  
There are  two immediate questions:   
(1) Is the correlation real?  (2) Is it smooth, or is there a roughly 
constant velocity for $r < 4$ kpc followed by a rapid increase beyond 
that point?  The issue is doubtful because the two uppermost points 
in the figure have much weaker statistical weight than those around 
$r \sim 2$ kpc.  A statistical assessment  suggests that  
the overall trend is {\it most likely\/} real, but the confidence level 
is only in the 75--80\% range,  and  the available data do not allow us to 
estimate the functional form.  For example, a linear fit for $r > 1$ kpc, 
and a very different two-velocity model with $V_{LSR} \sim$ constant out 
to 4 kpc, are about equally successful.\footnote{   
A standard weighted linear fit to $V_{LSR}(r)$ for $r > 1$ kpc 
has slope = $-1.9 \pm 1.7$ km s$^{-1}$ per kpc.  Hence the existence of  
 a positive slope is roughly an 88\%-confidence hypothesis. 
  We try three simple models: (a) a constant $V_{LSR} = -16.5$ km s$^{-1}$;  (b) the linear fit; 
   and (c) a two-value model with $V_{LSR} = -17.1$ km $^{-1}$ for $r < 4$ kpc and 
   $-5.9$ km s$^{-1}$ for $r > 4$ kpc.  The chi-squared sums (4.49, 3.29, and 1.73 
   respectively) indicate that all three models are plausible within the known statistical 
       errors, but model (c) fits best even if we allow for the fact that 
   it has three adjustable parameters.  Model (a) is noticeably 
   poorer by the same standard.  Intermediate functional forms give similar results. 
  Note that the data for Figure 10 is available on-line with the figure.}

We probably see through Hercules Cloud to its far side at $r \gtrsim 4$ kpc. 
The mean $V_{LSR}$ may or may not be increasing between 
$r \sim 1$ kpc and 4 kpc, but $V_{LSR}$ does appear to  increase at greater distances. 
This suggests that the rotational lag is less at larger distances, as our line of sight apparently passes through the Cloud to its far side,  as in the case of 
the star counts. 
The only caveat is that, at greater distances, our lines of sight are also at higher $|Z|$ distances
from the plane. For the Thick Disk stars in Q1, most of the stars at more than 4 kpc are 1.5 to
2 kpc above the plane. This trend is not as apparent for the MWTD with fewer stars per distance
bin and a higher velocity dispersion.  
This pattern for the Thick Disk corresponds very well with the over-density and star count excess discussed 
in Paper II.

With the addition of  velocities for fainter stars we have traced the kinematic asymmetry
and the enhanced rotational lag for the Thick Disk stars in Q1 out to 4 kpc or more.  
The rotational lag is greatest along our lines of sight toward the bar in the Disk.
At 3 to 4 kpc our sight lines extend above  the density contours mapped by the AGB stars on the
near side of the bar \citep{1992ApJ...384...81W}, but as our line of sight passes directly 
over the bar, 
the rotational lag appears to decrease. The kinematics are  consistent 
with a ``gravitational wake'' induced
by the rotating bar in the Disk that would trap and pile up stars {\it behind} 
it \citep{1992ApJ...400...80H,1998ApJ...493L...5D}.  Furthermore, the kinematic asymmetry 
and slower rotational lag in Q1 is shared by all three stellar
populations, the Thick Disk, MWTD, and the Disk. Interestingly, we find that the additional lag, 
relative to the same populations in Q4, is comparable for all three, at about 30 km s$^{-1}$.
The net  rotational lag in the direction of the bar for  the stellar populations  together with the 
over-density,  above and below the plane, support a dynamical interaction with the bar as the most 
likely explanation.
This conclusion does not necessarily eliminate the role of a merger remnant which could be interacting
with the bar. However,  the common stellar parameters, such as metallicity,  
demonstrate that the populations in Q1 and Q4 are the same, with asymmetric 
spatial and kinematic properties,  not two different populations.

Future work should include dynamical modeling of the interaction with the bar  and extended 
observations of more stars at the fainter magnitudes to trace the kinematics to greater 
distances across the bar.

\acknowledgments

This work was supported by  Collaborative National Science Foundation grants to 
 Humphreys (AST0507170), Cabanela (AST0729989), and Larsen (AST0507309).
They  thank the MMT Observatory, the support staff for the Hectospec and the 
telescope operators,  and NOAO and the staff at CTIO  for excellent observing 
support.  Beers and Lee  acknowledge partial funding of this work from grants
PHY 02-16783 and PHY 08-22648: Physics Frontier Center/Joint Institute
for Nuclear Astrophysics (JINA), awarded by the U.S. National Science
Foundation.

{\it Facilities:}  \facility{MMT/Hectospec, Bok/90Prime, CTIO/SMARTS/1.0m, CTIO/Blanco/Hydra}

\appendix
\section{Catalog of Velocities, Metallicities, and Atmospheric Parameters} 

The Catalog of velocities, metallicities, and the atmospheric parameters with their
erors are available as an on-line supplement to this paper. Table A1 includes all of the 
stars with V and B-V magnitudes and colors (Papers I and II). Table A2 is for those stars 
which only have photographic O and O-E magnitudes and colors  from the Minnesota Automated
Plates Scanner Catalog of the POSS I. It includes the stars from \citet{par04}. 
Examples are shown in Tables A1 and A2. The Catalog is also available at http://aps.umn.edu.  

\section{Mean V$_{LSR}$ and Rotation Rates for the Individual Fields in Q1 and Q4}    

\section{Mean V$_{LSR}$ and Rotation Rates for the Different Populations in the Individual Fields}

\clearpage

\begin{deluxetable}{lccccl}
\tabletypesize{\scriptsize}
\tablewidth{0pt}
\tablecaption{Observations}
\tablehead{
\colhead{Field}   & \colhead{$l$[$^{\circ}$]}  & \colhead{$b$ [$^{\circ}$]}
 &  Stars & \colhead{Total Exp Time}  & \colhead{Dates}\\
 &  & & \colhead{with velocities}  &  \colhead{[min]} & \colhead{Observed}
}
\startdata
CTIO/Hydra  2006\tablenotemark{a} & &  &  &  & \\[0.1in]
P802   &  339.5 & 33.0  &  96      &  120  & May 04, 05 \\
P855   &  309.0 & 37.0  &  73     &  120  & May 05     \\
P858   &  329.0 & 32.0  &  95     &  120  & May 04     \\
P910   &  303.0 & 30.0  &  71     &  120  & May 04     \\
P913   & 320.0  & 30.0  &  94     &  120  & May 05     \\
\hline                                                      
CTIO/Hydra  2008\tablenotemark{b}   & &  &  &  & \\[0.1in] 
H333+37  &  333.0   &   37.0   &      103    &   280 & Apr 08,10 \\
H330+20  &  330.0   &   20.0   &       96      &    240 & Apr 06,07,08,09,10\\
H327+40  &  327.0   &  40.0    &     105        &    250 & Apr 08,09 \\
H312+45  &  313.0   &  45.0    &      94      &    300 & Apr 07,09,10 \\
H310+31  &  310.0   &  31.0    &      64  &    300 & Apr 07,08,10 \\
H305+42  &  305.0   &  42.0    &      84   &    300 & Apr 06,08,09 \\
\hline                                                           
MMT/Hectospec\tablenotemark{b}  2007 \& 2008 & &   &  &  & \\[0.1in]   
H020+32    & 21.0  &  32.0  &  166   &    150  & Jun 13, 2007\\
H027+37     & 28.0   &  37.0    &     230   &    150  & May 31, 2008\\
H030+20   &  31.0    &  20.0    &     187    &    150  & Jun 15,19, 2007 \\
H030-20    & 31.0    &  -20.0   &     250    &    150  & Oct 17,2007,Sep 08,2008 \\
H033+40    & 33.0    &   40.0   &      229   & 150     & Jun 01, 2008 \\
H035+32     & 36.0   &   32.0   &      178   & 150     & May 14, Oct 14, 2007 \\  
H035-32     & 36.0   &   -32.0  &       184   & 150     & Jun 16,19 2007\\
H045-20     &  46.0  &   -20.0  &       210   &  150   &  Oct 18, 2007 \\
H048-45     & 49.0   &   -44.0  &       199   &  150   & Sep 09, 2008\\
H050+31    &  51.0   &    31.0  &       176   &  150   & Jun 13, Oct 18, 2007 \\
H050-31     & 51.0   &   -31.0  &       161   &  120   & Jun 18, 2007 \\
H060+20     & 61.0   &    20.0  &       209   &  90    &  Oct 15, 2007\\
H060-20     & 61.0   &   -20.0  &       215  &  150   &  Oct 14, 2007\\ 
\enddata

\tablenotetext{a}{Repeat of some fields observed by \citet{par04}. We use the same 
field identifications from the MAPS.}  
\tablenotetext{b}{Information for these fields is in \citep{lar2010a}.}

\end{deluxetable}

\clearpage
\begin{deluxetable}{lrcccc}
\tablewidth{0pt}
\tablecaption{Mean V$_{LSR}$ Velocities and Rotation Rates\tablenotemark{a}}
\tablehead{ 
\colhead{Quadrant} & \colhead{$N_{stars}$} & \colhead{{$<V_{LSR}>$}} & \colhead{$\sigma_{V_{LSR}}$} &  \colhead{$<\omega>$} & \colhead{$\sigma_{\omega}$}\\ 
  &    &  \colhead{km s$^{-1}$} &  \colhead{km s$^{-1}$}  & \colhead{km s$^{-1}$kpc$^{-1}$} & \colhead{km s$^{-1}$kpc$^{-1}$}
}
\startdata
Q1 above  & 1679   &   -28.6 $\pm$ 1.6 & 66.4   &  19.6 $\pm$ 0.5 & 21.0\\
Q1 below  & 1135    &   -35.6 $\pm$ 2.0\tablenotemark{b} & 67.4  &  19.7 $\pm$ 0.5\tablenotemark{b} & 15.3 \\
Q4 above  & 1170   &   -11.6 $\pm$ 2.0 & 66.7  &  31.3 $\pm$ 0.6 &  20.4 \\ 
\enddata
\tablenotetext{a}{Stars with V$_{LSR}$ greater than $\pm$ 200 km s$^{-1}$ are excluded
from the determination of the mean V$_{LSR}$ and mean $\omega$.}
\tablenotetext{b}{Without H048-45 (see text), $<V_{LSR}>$ and $<\omega>$ for Q1 below the plane are, 
respectively, $-$22.6 km s$^{-1}$ and 22.9 km s$^{-1}$kpc$^{-1}$.} 
\end{deluxetable}

\clearpage
\begin{deluxetable}{lllll}
\tablewidth{0pt}
\tablecaption{Criteria for Population Separation}
\tablehead{
\colhead{Population} &  \colhead{[Fe/H]}  & \colhead{[$\alpha$/Fe]} & \colhead{$|Z|$} & \colhead{r$_{\ast}$}
}
\startdata
Disk                 &     $\ge$ 0      &    \nodata    &   $\le$ 650 pc  &    \nodata  \\
                     &    0  to $-$0.3\tablenotemark{a}  &     \nodata &   $\le$ 650 pc  &    \nodata  \\
                     &    0  to  $-$0.5  &   $\le$ 0.2 &  $\le$ 650 pc &   \nodata \\   
                     &    0  to  $-$0.5  &   0.2 to 0.3 &  $\le$ 650 pc  &  \nodata \\  

Thick Disk           &    0  to $-$0.3\tablenotemark{a}  &     \nodata &  $>$ 650 pc &   $\le$ 5 kpc\\ 
                     &   $-$0.3 to $-$0.5\tablenotemark{a} &     \nodata & \nodata  & $\le$ 5 kpc \\
                     &    0  to  $-$0.5  &   $<$ 0.3 &  $>$ 650 pc &   $\le$ 5 kpc\\ 
                     &    0  to  $-$0.5  &   $>$ 0.3    &   \nodata  & $\le$ 5 kpc \\
                     &    $<$ $-$0.5 to $-$0.8 &   \nodata    &   \nodata  &   $\le$ 5 kpc\\
		     &    $-$0.8 to $-$1.2 &   \nodata    &  $<$ 2 kpc &   $\le$ 5 kpc\\

MWTD                 &    $<$ $-$0.8 to $-$1.2 & \nodata    & $\le$ 2 $|Z|$ $\le$ 4 kpc  &  \nodata\\
                     &    $-$1.2 to $-$1.8  & \nodata    & $<$  4 kpc  &  \nodata \\

Halo                 &    $-$1.2 to $-$1.8  & \nodata    & $\ge$ 4 kpc &  \nodata \\
                     &    $<$ $-$1.8  &   \nodata    &   \nodata  &  \nodata \\
\enddata
\tablenotetext{a}{Used for stars without an [$\alpha$/Fe] estimate. See text} 
\end{deluxetable}

\clearpage
\begin{deluxetable}{llccccc}
\tabletypesize{\scriptsize}
\tablewidth{0pt}
\tablecaption{Mean [Fe/H], LSR Velocities and Rotation Rates for the Four Populations}
\tablehead{
\colhead{Population} & \colhead{$N_{stars}$} & \colhead{[Fe/H]} & \colhead{$<V_{LSR}>$}  
& \colhead{$\sigma_{V_{LSR}}$} &  \colhead{$<\omega>$}  & \colhead{$\sigma_{\omega}$}\\
& & & \colhead{km s$^{-1}$} &  \colhead{km s$^{-1}$} &  \colhead{km s$^{-1}$kpc$^{-1}$} 
& \colhead{km s$^{-1}$kpc$^{-1}$} }
\startdata
Disk     &       &       &       &       &       &              \\[0.1in]
Q1A    & 39   & $-$0.3 &  $-$14.0 $\pm$ 5.5 & 34.2 & 24.4 $\pm$ 1.2  & 7.2    \\
Q1B    &  105  &  $-$0.2 &  0.3 $\pm$ 3.5 & 35.9 & 27.6 $\pm$ 0.8  & 7.7    \\  
Q4         &  70 &  0.0     &  $-$19.9 $\pm$ 4.1 &  34.5 & 32.3 $\pm$ 1.0 & 8.7 \\ 
\hline                                                        \\
Thick Disk      &      &       &       &       &       &              \\[0.1in]
Q1A     &  793  &  $-$0.8 &  $-$-18.4 $\pm$ 1.9 & 52.2 &  23.2 $\pm$ 0.5 &   15.3 \\ 
Q1B    &  453  &  $-$0.7  & $-$12.5 $\pm$  2.3 & 49.4 &  25.0 $\pm$ 0.5 &  10.4 \\ 
Q4         &  555 &   $-$0.7  & $ -$20.1 $\pm$ 2.3 & 53.5 &  33.5 $\pm$ 0.7 & 15.6  \\  
\hline
MWTD &      &       &       &       &       &              \\[0.1in]
Q1A     &  379  & $-$1.3 & $-$25.2 $\pm$ 3.5 &  61.6 &   20.1 $\pm$ 1.0 & 19.9  \\      
Q1B      & 117  & $-$1.4 & $-$-41.0 $\pm$ 5.7 &  61.5 &   19.2 $\pm$ 1.2 & 13.2 \\   
Q4           & 216  & $-$1.3 &  $-$10.2 $\pm$  4.5  & 65.7  & 31.6 $\pm$ 1.4 &   20.7 \\ 
\hline                                                                  
Halo   &      &       &       &       &       &              \\[0.1in]
Q1A       &  337  & $-$1.8  & $-$82.7 $\pm$ 4.9 & 90.3 &   3.0 $\pm$ 1.6 &  29.4 \\   
Q1B      &   108  & $-$1.9 & $-$88.9 $\pm$ 9.7 &  100.7 &   9.5 $\pm$ 2.2 & 22.7\\ 
Q4           &  103   & $-$1.9 &  48.0   $\pm$11.3  & 115.0  &  15.3  $\pm$ 3.6 &   36.3 \\  
\hline
High Velocity\tablenotemark{a} &    &    &       &       &       &              \\[0.1in]   
Q1A  &  117  &  -1.6  &  -238.5 $\pm$   11.0 &  118.6 &  -35.2  $\pm$ 4.0 &   43.4\\ 
Q1B  &  66   &  -1.6  &  -281.8 $\pm$   6.2  & 50.6   &  -30.2 $\pm$  1.9  &  15.6\\ 
Q4       &  29   &   -1.4 &   209.1 $\pm$  34.1 &  183.8 &  -25.1 $\pm$    9.1 &   48.9\\                                                                        
\enddata
\tablenotetext{a}{The high velocity stars (V$_{LSR}$ $\ge$ $\pm$ 200 km s$^{-1}$) were not included 
in the four populations. They are listed separately here although, they are presumably members of the Halo.}
\end{deluxetable}

\clearpage

\begin{deluxetable}{lrcccc}
\tabletypesize{\scriptsize}
\tablewidth{0pt}
\tablecaption{Solutions for the Velocity Components}
\tablehead{
\colhead{Population} & \colhead{$N_{stars}$} & \colhead{$v_{r}$} &  
\colhead{$v_{\phi}$} &
 \colhead{$v_{z}$} & \colhead{$\sigma$} \\
 &     &  \colhead{km s$^{-1}$} & \colhead{km s$^{-1}$} & 
   \colhead{km s$^{-1}$} &
  \colhead{km s$^{-1}$}
}
\startdata
Disk  &       &       &       &       &                     \\[0.1in]  
Q1  &  145  &    0 $\pm$ 10 &  193 $\pm$ 8  &  -19 $\pm$ 8  &  35  \\ 
    &       &   -3 $\pm$ 10 &  194 $\pm$ 8  &       (0)     &  36  \\  
    &       &       (0)     &  196 $\pm$ 4  &       (0)     &  36  \\          
Q1A &   40  &  +12 $\pm$ 50 &  194 $\pm$ 15 &  -7 $\pm$ 57  &  36  \\ 
    &       &  +18 $\pm$ 20 &  193 $\pm$ 14 &      (0)      &  36  \\  
    &       &       (0)     &  183 $\pm$  8 &      (0)      &  36  \\  
Q1B &  105  & -177 $\pm$ 97 & 408 $\pm$ 118 &  +721 $\pm$ 405 & 35  \\  
    &       &   -5 $\pm$ 11 & 198 $\pm$  9  &      (0)      &  35  \\  
    &       &       (0)     & 202 $\pm$  5  &      (0)      &  35  \\  
Q4  &   69  &  +34 $\pm$ 29 & 220 $\pm$ 34  &  +18 $\pm$ 57 &  35  \\  
    &       &  +27 $\pm$ 19 & 211 $\pm$ 20  &      (0)     & 35  \\  
    &       &       (0)     & 237 $\pm$  7  &      (0)     & 35  \\       
\hline                         \\  
Thick Disk  &   &   &   &   &  \\[0.1in]
Q1  & 1245  &   -2 $\pm$  5 &  161 $\pm$ 4  &  -14 $\pm$ 3 & 50  \\
    &       &   +7 $\pm$  5 &  165 $\pm$ 4  &     (0)      & 51 \\
    &       &       (0)     &  161 $\pm$ 2  &     (0)      & 51 \\
Q1A &  793  &  -27 $\pm$ 11 &  169 $\pm$ 7  & -50 $\pm$ 15 & 51 \\
    &       &   +4 $\pm$ 6  &  155 $\pm$ 5  &      (0)     & 51 \\
    &       &        (0)    &  152 $\pm$ 3  &      (0)     & 51 \\
Q1B &  452  & -32 $\pm$ 12  &  196 $\pm$ 9  & +84 $\pm$ 24 & 49 \\
    &       &  -3 $\pm$ 9   &  169 $\pm$ 5  &      (0)     & 49\\
    &       &      (0)      &  171 $\pm$ 3  &      (0)     &  50 \\
Q4  & 551   &  +43 $\pm$ 11  &   182 $\pm$ 12 & +2 $\pm$ 16 &  54 \\
    &       &  +42 $\pm$ 9   &   181 $\pm$ 8  &  (0)    &  53\\
    &       &       (0)      &   217 $\pm$ 4  &  (0)    &  54 \\
\hline                                                     \\
MWTD  &       &       &       &       &                     \\[0.1in]
Q1  &  492  & -13 $\pm$ 12 &  113 $\pm$ 7  & -8 $\pm$ 7  & 63 \\
    &       &   -7 $\pm$ 10 & 115 $\pm$ 7     &  (0)    & 63\\
    &       &   (0)      & 119 $\pm$ 4    &   (0)    & 63\\
Q1A & 377   &   -24 $\pm$ 19 & 136 $\pm$ 14   & -43 $\pm$ 25 & 64 \\
    &       &  +2 $\pm$ 12   & 120 $\pm$ 10   &  (0)    & 64 \\
    &       &   (0)      & 118 $\pm$ 5    &  (0)    & 64 \\
Q1B & 115   &  -46 $\pm$ 26  & 103 $\pm$ 30   & +5 $\pm$ 56 & 60\\
    &       &  -45 $\pm$ 22  & 100 $\pm$ 12   &  (0)    & 60\\
    &       &   (0)      & 120 $\pm$ 7    &  (0)    & 61 \\
Q4  &  216  &  -1 $\pm$ 22   & 138 $\pm$ 24   &  -49 $\pm$ 36 & 64 \\
    &       &  +21 $\pm$ 15  & 165 $\pm$ 13   & (0)    & 64 \\
    &       &    (0)      & 181 $\pm$ 7   & (0)    & 64 \\
\enddata
\end{deluxetable}

\begin{landscape}
\begin{deluxetable}{lllllllclllll}
\tabletypesize{\tiny}
\tablenum{A1}
\tablecaption{Catalog of Velocities, Metallicity, and Atmospheric Parameters (CCD Photometry)}
\tablehead{
 \colhead{Object} & \colhead{{\it l}} & \colhead{{\it b}} & \colhead{V} & \colhead{$B-V$}
 &  \colhead{$V_{Hel}$} & \colhead{$V_{LSR}$} & \colhead{$\omega$} & \colhead{[Fe/H]} & \colhead{[$\alpha$/Fe]} & \colhead{$T_{eff}$} & \colhead{log{\it g}} & Dist. \\
                   &  \colhead{deg}    &  \colhead{deg}   &  \colhead{mag}  &  \colhead{mag}   &  \colhead{km s$^{-1}$}  &   \colhead{km s$^{-1}$}  &  \colhead{km s$^{-1}$ kpc$^{-1}$}  &   &  & $^{\arcdeg}$K   &    &  \colhead{kpc} \\
}
\startdata
 H060+20-022532 &  60.54 & +19.99 & 18.68 & +0.59 & -56.6(4.1) & -40.2 & +21.4 & -1.01(0.02) & +0.24(0.04) &  5691(34) & 3.87(0.35) &  3.0  \\ 
 H060+20-020900 &  60.53 & +20.03 & 18.39 & +0.60 &  +2.4(4.6) & +18.8 & +30.4 & -0.43(0.02) & +0.18(0.04) &  5683(82) & 4.36(0.28) &  2.3  \\ 
 H060+20-018959 &  60.50 & +20.07 & 17.10 & +0.73 & -400.3(2.7) & -383.9 & -31.2 & -1.01(0.03) & +0.30(0.04) &  5702(61) & 4.19(0.40) &  2.2  \\ 
 H060+20-018109 &  60.48 & +20.22 & 16.30 & +0.63 & -322.2(3.6) & -305.9 & -19.3 & -1.03(0.02) & +0.02(0.06) &  5896(39) & 3.97(0.25) &  4.0  \\ 
 H060+20-015828 &  60.47 & +20.29 & 16.82 & +0.74 & +16.6(1.7) & +32.9 & +32.5 & -2.19(0.05) & +0.08(0.06) &  5341(40) & 3.02(0.08) &  3.7  \\ 
 H060+20-023202 &  60.46 & +19.95 & 18.16 & +0.73 & +43.0(3.0) & +59.4 & +36.6 & -0.70(0.05) & +0.12(0.03) &  5720(53) & 3.70(0.43) &  1.0  \\ 
\enddata
\end{deluxetable}
\end{landscape}

\begin{landscape}
\begin{deluxetable}{lllllllclllll}
\tabletypesize{\tiny}
\tablenum{A2}
\tablecaption{Catalog of Velocities, Metallicity, and Atmospheric Parameters (MAPS Photometry)}
\tablehead{
 \colhead{Object\tablenotemark{a}} & \colhead{{\it l}} & \colhead{{\it b}} & \colhead{O} & \colhead{$O-E$}
 &  \colhead{$V_{Hel}$} & \colhead{$V_{LSR}$} & \colhead{$\omega$} & \colhead{[Fe/H]} & \colhead{[$\alpha$/Fe]} & \colhead{$T_{eff}$} & \colhead{log{\it g}} & Dist. \\
                  &  \colhead{deg}    &  \colhead{deg}   &  \colhead{mag}  &  \colhead{mag}   &  \colhead{km s$^{-1}$}  &   \colhead{km s$^{-1}$}  &  \colhead{km s$^{-1}$ kpc$^{-1}$}  &   &  & $^{\arcdeg}$K   &    &  \colhead{kpc} \\
}
\startdata
P332-1931756 &  51.26 & +37.07 & 15.89 & +0.43 & +12.5(3.6) & +28.6 & +33.2 & -0.09(0.08) & -0.01(0.05) &  5430(31) & 3.90(0.56) &  2.4  \\ 
P332-1925176 &  51.21 & +37.31 & 15.56 & +0.57 & -66.4(5.5) & -50.3 & +17.4 & -0.74(0.03) & +0.47(0.02) &  5349(19) & 4.88(0.17) &  1.3  \\ 
P332-1922793 &  51.12 & +37.37 & 15.79 & +0.35 & -84.8(9.9) & -68.7 & +13.6 & -0.34(0.02) & +0.26(0.03) &  5596(29) & 4.08(0.01) &  3.7  \\ 
P332-1923351 &  51.11 & +37.35 & 15.98 & +0.47 & -72.8(2.9) & -56.7 & +16.1 & -1.20(0.10) & +0.51(0.03) &  5343(51) & 4.53(0.06) &  2.2  \\ 
P332-1925685 &  51.09 & +37.26 & 16.12 & +0.53 & -55.6(1.7) & -39.5 & +19.5 & -0.63(0.04) & +0.38(0.02) &  5455(80) & 4.88(0.11) &  1.8  \\ 
P332-2071813 &  51.04 & +36.98 & 15.77 & +0.46 & -32.6(2.9) & -16.4 & +24.2 & -2.53(0.18) & +0.11(0.08) &  6370(38) & 3.07(0.36) &  2.3  \\ 
\enddata
\tablenotetext{a}{The stars in this table are from the Minnesota Automated Plate Scanner (MAPS) Catalog
of the POSS I. Their complete designation is ``MAPS-PXXX-YYYYYY''.}  
\end{deluxetable}
\end{landscape}

\begin{deluxetable}{llrcccc}
\tabletypesize{\scriptsize}
\tablenum{B}
\tablewidth{0pt}
\tablecaption{Mean LSR Velocities and Rotation Rates for  Individual Fields in Q1 and Q4\tablenotemark{a}} 
\tablehead{
\colhead{Field} & \colhead{{\it l, b}} & \colhead{$N_{stars}$} & \colhead{$<V_{LSR}>$}  
& \colhead{$\sigma_{V_{LSR}}$} &  \colhead{$<\omega>$}  & \colhead{$\sigma_{\omega}$}\\
& & & \colhead{km s$^{-1}$} &  \colhead{km s$^{-1}$} &  \colhead{km s$^{-1}$kpc$^{-1}$} & 
\colhead{km s$^{-1}$kpc$^{-1}$} 
}
\startdata
Q1 above  &  &  &  & &  & \\[0.1in]
H060+20     & 61.0, 20.0  & 193   &  -33.9 $\pm$ 3.8  & 52.9  & 22.3 $\pm$ 0.6 & 8.4\\      
P332 &  51.0, 37.0 & 37 & -45.8 $\pm$ 9.2 & 56.0 &  18.2 $\pm$ 1.9 & 11.3\\
H050+31    &  51.0, 31.0  & 157  &  -50.6  $\pm$ 4.2 & 55.9 & 17.9 $\pm$ 0.8 & 10.6\\ 
P387 &  40.0, 41.0 &  57 & -35.0 $\pm$ 7.6 & 57.8 & 18.4 $\pm$ 2.0 & 15.0\\
P448 &  40.0, 30.0 &  60 & -15.9 $\pm$ 8.1 & 63.4 & 23.9 $\pm$ 1.8 & 14.2\\
H035+32   & 36.0, 32.0  & 165 &  -32.1 $\pm$ 5.2  & 67.3  &  19.3 $\pm$ 1.4& 17.3 \\  
H033+40    & 33.0, 40.0  & 217&  -31.8 $\pm$ 4.2& 62.5  &  17.8 $\pm$ 1.3 & 19.0 \\ 
P507 &  31.0, 32.0 &  65 & -20.7 $\pm$ 8.6 & 69.5 & 21.6 $\pm$ 2.5 & 20.0\\
H030+20   &  31.0, 20.0 & 174 & -12.6 $\pm$ 5.9  & 78.4 &  24.2 $\pm$ 1.6  & 20.9 \\ 
H027+37     & 28.0, 37.0& 219  & -25.3 $\pm$ 4.8 & 70.9  &  18.8 $\pm$ 1.7 & 24.4\\ 
P505 &  21.0, 42.5 &  62 & -26.2 $\pm$ 8.5 & 67.2 & 15.0 $\pm$ 4.0 & 32.0\\
H020+32    & 21.0, 32.0  & 273 &  -20.9 $\pm$ 4.3 & 70.2  & 18.5 $\pm$ 1.8 & 30.2\\ 
\hline\\
Q1 below  &  &  &  & &  & \\[0.1in]
H060-20     & 61.0, -20.0  & 202  &  -29.8 $\pm$ 3.8  & 54.4 & 22.9 $\pm$ 0.6 & 8.1\\ 
H050-31     & 51.0, -31.0  & 139  &  -52.5  $\pm$ 6.1 & 72.1  & 17.5 $\pm$ 1.2 & 13.7\\ 
H048-45     & 49.0, -44.0  & 182  &  -104.1 $\pm$ 3.5 & 46.6  &  2.5 $\pm$ 0.8& 11.2\\ 
H045-20     &  46.0, -20.0  & 198 &  -8.9  $\pm$ 4.2& 58.7  & 25.8 $\pm$ 0.8& 11.0\\  
H035-32     & 36.0, -32.0  & 169  &  -23.9  $\pm$ 5.3 & 69.0 & 21.3 $\pm$1.4 & 17.8\\  
H030-20    & 31.0, -20.0   & 245  &  -9.8  $\pm$ 3.5 & 55.3 & 24.9 $\pm$ 0.9 & 14.7\\  
\hline\\ 
Q4 above  &  &  &  & &  & \\[0.1in]
P910   &  303.0, 30.0  & 69  &   -0.9 $\pm$ 6.7 & 55.4 & 27.6 $\pm$ 1.2 & 9.6 \\
H305+42  &  305.0, 42.0 & 80 &   5.0  $\pm$ 6.3 & 56.7  & 26.5 $\pm$ 1.3& 12.5\\  
P855   &  309.0, 37.0  & 78 &   3.9  $\pm$ 6.2  & 54.8  & 26.7 $\pm$ 1.3& 11.1\\  
H310+31  &  310.0, 31.0 & 64 &   8.7  $\pm$ 8.2 & 65.5  & 25.8 $\pm$ 1.6 & 12.5\\ 
H312+45  &  313.0, 45.0 & 93 &  -5.4  $\pm$ 6.1& 58.8  & 28.8 $\pm$ 1.5 & 14.1\\ 
P799    &  320.0,  41.0 & 60 &  -8.8  $\pm$ 8.6& 66.6  & 29.8 $\pm$ 2.2 & 17.2\\ 
P913   & 320.0, 30.0  & 120 &  -22.4 $\pm$ 4.9 & 53.3  & 32.5 $\pm$ 1.1& 12.0\\  
H327+40  &  327.0, 40.0 & 103 & -6.3 $\pm$ 7.3& 74.5  & 29.4 $\pm$ 2.2 & 22.3\\  
P858   &  329.0, 32.0  & 114 & -9.4 $\pm$ 6.1& 65.5   & 30.2 $\pm$ 1.8 & 18.9\\ 
H330+20  &  330.0, 20.0 & 93 & -40.4 $\pm$ 7.3 & 70.5  &  38.3 $\pm$2.0 & 18.8\\  
H333+37  &  333.0, 37.0 & 101 & 1.6   $\pm$ 7.4& 74.7 &  26.9 $\pm$ 2.6 & 25.7\\ 
P741   &  339.0, 41.0 & 63   &  -33.3 $\pm$ 8.3& 66.0 &  42.8 $\pm$ 3.8 & 30.3\\ 
P802   &  339.5, 33.0  & 122 & -28.5 $\pm$ 5.7& 63.1 &  39.5 $\pm$ 2.4 & 26.8\\ 
\enddata
\tablenotetext{a}{Stars with V$_{LSR}$ greater than $\pm$ 200 km s$^{-1}$ are excluded
from the determination of the mean V$_{LSR}$ and mean $\omega$.}
\end{deluxetable}

\begin{deluxetable}{lrccccc}
\tabletypesize{\scriptsize}
\tablenum{C}
\tablewidth{0pt}
\tablecaption{Kinematics for the Populations in the Individual Fields}
\tablehead{
\colhead{Field} & \colhead{{\it l/b}} & \colhead{$N_{stars}$} & \colhead{$<V_{LSR}>$}
& \colhead{$\sigma_{V_{LSR}}$} &  \colhead{$<\omega>$}  & \colhead{$\sigma_{\omega}$}\\
& & & \colhead{km s$^{-1}$} &  \colhead{km s$^{-1}$} &  \colhead{km s$^{-1}$kpc$^{-1}$}
& \colhead{km s$^{-1}$kpc$^{-1}$} }
\startdata
Q1 above    &        &      &        &          &           &        \\[0.1in]  
\hline                                                        \\
H060+20     &  61/+20 &      &        &          &           &        \\   
  Disk      &         &   21 &   -11.4 $\pm$ 8.2  &  37.8  &  25.7 $\pm$ 1.3  & 5.8\\ 
  ThDi      &         &   118 &  -33.2 $\pm$  4.6 &   49.9 &   22.4 $\pm$  0.7 &    7.7\\ 
  MWTD      &         &   24  &  -29.3 $\pm$ 10.4 &   50.9 &   23.0$\pm$ 1.6  &   7.8\\  
  Halo      &         &   16  &   -59.0 $\pm$ 19.0 &  76.0 &   18.4$\pm$  2.9 &   11.7\\   
P332        & 51/+37  &       &        &          &           &        \\   
 ThDi       &         &   27  & -42.1 $\pm$  9.6  &  49.9  &  18.9 $\pm$ 1.9 &   10.1\\  
 MWTD       &         &    4  &  -2.4 $\pm$ 18.5  &  37.1  &  27.0 $\pm$ 3.8 &    7.6\\   
 Halo       &         &  3    & -110.8 $\pm$  48.7 &   84.4 &  5.1 $\pm$  9.9 &   17.1\\
H050+31     & 50/+31  &      &        &          &           &        \\  
 Disk       &         &    2 &   -47.4 $\pm$ 27.0 &   38.1 &   18.5 $\pm$ 5.1 &    7.3\\   
 ThDi       &         &   122&   -35.8 $\pm$  4.0 &   44.2 &   20.7 $\pm$  0.8 &    8.4\\  
 MWTD       &         &   27 &   -47.0 $\pm$ 9.2  &  47.6  &  18.6 $\pm$  1.7 &    9.0\\   
 Halo       &         &   29 & -128.6 $\pm$ 10.4  &  56.3  &   3.1  $\pm$ 2.0 &   10.7\\  
P387        &  40/+41 &      &        &          &           &        \\     
 ThDi       &        &    25 &  -22.7 $\pm$ 9.6 &   47.8   &  21.6 $\pm$2.5 &   12.3\\  
 MWTD       &        &    23 &  -39.5 $\pm$ 9.6 &   45.8   &   17.3 $\pm$ 2.5 & 11.83\\   
 Halo       &        &    9  &  -75.3 $\pm$ 33.1 &   99.3  &   7.8 $\pm$  8.6 &   25.9 \\    
P448        & 40/+30 &       &         &         &           &         \\     
 ThDi       &        &    17 &  -20.0 $\pm$  13.7&    56.6 &   23.0 $\pm$ 3.1 &   12.7\\    
 MWTD       &        &    22 &  -13.3 $\pm$ 15.3 &   71.6  &  24.5 $\pm$  3.4 &   16.1\\    
 Halo       &        &    12 &  -28.6 $\pm$ 20.8 &   72.2  &  21.0 $\pm$  4.7 &   16.2\\   
H035+32     &  36/+32  &      &        &          &           &        \\     
 ThDi       &        &    95&    -8.0 $\pm$ 4.9 &   48.1  &  25.4 $\pm$  1.3  &  12.4\\ 
 MWTD       &        &   39 &  -28.1 $\pm$  9.9 &   61.8  &  20.3 $\pm$  2.5  &  15.9\\   
 Halo       &        &  31  &  -124.7 $\pm$ 11.0 &   61.5 & -4.5  $\pm$  2.8  &  15.8\\   
H033+49     & 33/+40 &      &          &           &          &         \\     
 Disk       &        &   10 &   -16.1$\pm$ 10.8 &   34.3  &  22.6 $\pm$ 3.3 &   10.5\\   
 ThDi       &        &   68 &  -24.9$\pm$  5.5  &  45.2  &  19.9 $\pm$ 1.7  &  13.8\\  
 MWTD       &        &   48 &  -18.8 $\pm$ 8.1 &   56.1  &  21.8 $\pm$ 2.5  &  17.1 \\    
 Halo       &        &   65 &  -82.5$\pm$ 10.2 &   82.0  &   2.5 $\pm$ 3.1  &  24.9 \\   
P507        & 31/+32       &      &                  &         &                  &   \\    
 ThDi       &        &   22 &   -6.4 $\pm$ 10.7 &   50.4  &  25.7 $\pm$ 3.1 &   14.5\\   
 MWTD       &        &   23 &  -36.6 $\pm$ 17.0 &   81.7  &  16.9 $\pm$ 4.9 &   23.4\\  
 Halo       &        &   10 &  -82.2 $\pm$ 28.9 &   91.3  &   3.9 $\pm$ 8.3 &   26.3\\   
H030+20     & 30/+20 &      &                &       &                       &\\   
 Disk       &        &    3 & -2.6 $\pm$ 2.8 &    4.8 &   26.8 $\pm$ 0.7 &   1.3\\    
 ThDi       &        &   86 & 8.5 $\pm$ 5.7  &   53.0 &   29.8 $\pm$ 1.5 &   14.1\\   
 MWTD       &        &    36& -19.0 $\pm$ 11.4 &   68.2 &   22.5 $\pm$ 3.0 &   18.2\\   
 Halo       &        &    36& -83.4 $\pm$ 19.7 &  118.4 &  5.2 $\pm$  5.3 &   31.6\\  
H027+37     & 28/+37 &      &                  &         &                &      \\   
 Disk       &        &    3 &  -14.1 $\pm$ 12.6 &   21.8 &   22.7 $\pm$ 4.2 &    7.3\\  
 ThDi       &        &   86 &   -4.7 $\pm$ 5.7  &  52.6  &  25.9 $\pm$ 1.9  &  18.1 \\    
 MWTD       &        &  50  & -24.4 $\pm$ 8.8   &  62.2  &  19.1 $\pm$ 3.0  &  21.4\\   
 Halo       &        &  63  &  -74.1 $\pm$ 11.8 & 93.4   &  2.0 $\pm$ 4.0  &  32.1\\  
P505        & 21/+42.5 &    &                   &        &                 &      \\   
 ThDi       &        &   19 &  -18.9 $\pm$ 13.2 &   57.6 &   18.6 $\pm$ 6.2 &   27.1\\   
 MWTD       &        & 27   &  -22.6 $\pm$ 12.2 &   63.3 &   16.7 $\pm$ 5.8 &   30.0 \\  
 Halo       &        &  14  & -52.7 $\pm$ 26.0  &  97.5  &   2.1 $\pm$ 12.5 &   46.9\\   
H020+32     & 21/+32 &      &                   &        &                  &        \\    
 ThDi       &        &  108 &  -15.0 $\pm$ 5.4  &  56.5  &  21.0 $\pm$  2.3 &   24.3\\ 
 MWTD       &        &   56 &  -18.0 $\pm$ 8.6  &  64.3  &  19.7 $\pm$ 3.7  &  27.7\\    
 Halo       &        &   49 &  -68.7 $\pm$ 13.8 &   96.7 &   -2.1 $\pm$ 6.0 &   41.8\\   
\hline                                                        \\
Q1 below   &        &      &        &          &           &        \\[0.1in]
\hline                                                        \\
H060-20    & 61/-20 &      &        &          &           &         \\    
 Disk      &        &    36&  -9.7 $\pm$ 5.5   & 33.0 &   26.0 $\pm$ 0.8  &   5.1\\    
 ThDi      &        &   124& -21.8 $\pm$ 4.2 &   47.1 &   24.1 $\pm$ 0.6  &   7.2 \\    
 MWTD      &        &    12&   -65.8 $\pm$13.6  &  47.1 &  17.4$\pm$ 2.1 &    7.2 \\ 
 Halo      &        &    14&  -116.3 $\pm$ 21.8 &   81.8 &  9.7 $\pm$3.4 &   12.6\\   
H050-31    & 51/-30 &  &        &          &           &         \\ 
 ThDi      &        &  55  &  -19.0 $\pm$ 6.3  &  46.7 &   23.9 $\pm$  1.2 &    8.9\\ 
 MWTD      &        &  42  & -50.4 $\pm$ 9.6 &   61.9  &  17.9 $\pm$ 1.8 &   11.8\\  
 Halo      &        &  33  &  -111.0$\pm$  15.4 &   88.3  & 6.4 $\pm$ 2.9 &   16.8 \\   
H048-45    & 49/-44 &      &                    &         &                &      \\   
 Disk      &        &   1  & -70.5              & \nodata  &    10.6     & \nodata \\  
 ThDi      &        &  91  &  -89.2 $\pm$  3.9 &   37.6  &   6.1 $\pm$  0.9  &   9.0\\       
 MWTD      &        &  16  &  -86.0$\pm$ 9.5  &  38.0   &  6.9 $\pm$ 2.3 &    9.1\\  
 Halo      &        &  42  & -159.1$\pm$ 5.9  &  38.2   & -10.8$\pm$ 1.4 &    9.2\\   
H045-20    & 45/-20 &      &                  &         &                &        \\   
 Disk      &        & 26   &  16.3$\pm$ 7.0   &  35.7   & 30.5 $\pm$ 1.3 &    6.7 \\   
 ThDi      &        & 132  & -1.4 $\pm$ 4.3   & 48.9   &  27.2 $\pm$ 0.8 &    9.2\\   
 MWTD      &        & 15   & -44.3$\pm$ 16.3  &  63.0  &  19.2$\pm$ 3.1  &  11.8\\    
 Halo      &        &  18  & -94.6$\pm$ 18.4  &  78.0  &   9.7$\pm$ 3.5  &  14.7\\  
H035-32    & 35/-32 &      &                  &        &                 &      \\    
 ThDi      &        & 95   & -17.1 $\pm$ 5.5  &  53.3  &  23.1 $\pm$ 1.4 &   13.7\\    
 MWTD      &        & 40   &  -23.5$\pm$10.6  &  66.9  &  21.4 $\pm$ 2.7 &   17.3\\  
 Halo      &        & 31   & -58.8 $\pm$20.2  &  112.7 &  12.4 $\pm$ 5.2 &   29.1\\     
H030-20    & 31/-20 &      &                  &        &                 &      \\    
 Disk      &        &  43  & -1.0$\pm$ 5.5    &  35.8  &  27.2 $\pm$ 1.5 &  9.5\\  
 ThDi      &        &  48  & -5.3$\pm$ 7.2    & 49.7   &  26.1 $\pm$ 1.9 &   13.3\\    
 MWTD      &        &  8   & -35.7$\pm$ 7.2   & 20.4   & 18.0$\pm$ 1.9   &  5.4\\    
 Halo      &        &  12  & -65.2$\pm$ 39.1  & 135.4  &  10.2$\pm$ 10.4 &  35.9\\   
\hline                                                        \\
Q4   &        &      &        &          &           &        \\[0.1in]
\hline                                                        \\
P910     & 303/+30  &         &          &           &       & \\    
 Disk    &          &  11  & -12.8 $\pm$  11.1 & 36.8 &  29.7$\pm$ 1.9  &   6.4\\    
 ThDi    &          &  43  &  -6.9 $\pm$  6.7  & 43.9 &  28.7$\pm$ 1.2  &   7.6 \\    
 MWTD    &          &   9  &  12.4$\pm$  20.8  & 62.5 & 25.3$\pm$  3.6  &  10.8\\    
 Halo   &           &   3  & 115.4$\pm$  73.2  & 126.8&  7.5$\pm$ 12.7  &  21.9\\   
H305+42  & 305/+42 &       &                   &      &                 &      \\     
 ThDi    &         &  40   & -14.0$\pm$ 7.0 &   44.4  &  30.4 $\pm$ 1.4 &  9.1\\   
 MWTD    &         &   8   &  17.3$\pm$ 22.4 & 63.3   & 24.0 $\pm$ 4.6  &  13.0\\    
 Halo    &         &  11   & 90.6 $\pm$ 24.9 & 82.5   &  8.8 $\pm$ 5.1  &  17.0\\    
P855     & 309/+37 &       &                   &      &                 &      \\     
 Disk    &         &   9   & -6.3$\pm$ 7.4 &   22.3  &  28.8$\pm$ 1.5   &  4.5 \\    
 ThDi    &         &  39   & -2.6$\pm$ 9.0 &   56.3  &  28.0$\pm$ 1.8   & 11.4 \\   
 MWTD    &         &  6    & 44.3$\pm$ 29.5 & 72.3   & 18.5$\pm$ 6.0    & 14.6 \\   
 Halo    &         &   5   &  73.9$\pm$25.7 & 57.4   & 12.6$\pm$ 5.2    & 11.6 \\    
H310+31  & 310/+51 &       &                   &      &                 &      \\   
 ThDi    &         &  39   &  -7.2$\pm$ 8.6 & 53.5  &  28.9 $\pm$ 1.6   & 10.2  \\    
 MWTD    &         &  16   & 2.2$\pm$ 17.3  &  69.4 &   27.1$\pm$ 3.3   & 13.3 \\   
 Halo    &         &   4   & 127.1$\pm$ 26.9 & 53.9 &  3.3 $\pm$ 5.1    & 10.2\\    
H312+45  & 313/+45 &       &                   &      &                 &      \\   
 Disk    &         &  1    &-37.9            & \nodata &   27.4         & \nodata \\  
 ThDi    &         & 38    & -19.9$\pm$ 6.1  &  37.8   &  31.3 $\pm$ 1.5 &  9.1 \\   
 MWTD    &         & 17    & -13.0$\pm$16.6  &  68.5   & 30.6$\pm$  4.0  &  16.4 \\   
 Halo    &         & 11    & 44.2$\pm$ 19.7  &  65.2   & 16.9$\pm$  4.7  &  15.7 \\    
P799     & 320/+41 &       &                   &      &                 &      \\   
 ThDi    &         & 11    & -2.7$\pm$ 23.3  &  77.2  &  28.3$\pm$ 6.0  &  19.9\\   
 MWTD    &         & 12    & -15.0$\pm$ 13.9 &  48.0  &  31.4$\pm$ 3.6  &  12.4 \\  
 Halo    &         &  2    & -37.2 $\pm$ 2.4 &  3.3   & 37.1 $\pm$ 0.6  &   0.8\\   
P913     & 320/+30 &       &                   &      &                 &      \\ 
 Disk    &         & 22    & -28.5$\pm$ 5.6  &  26.2  &  33.9$\pm$ 1.3  &   5.9 \\   
 ThDi    &         & 65    & -24.2$\pm$ 6.6  &  53.3  &  32.9$\pm$ 1.5  &  12.0 \\   
 MWTD    &         & 12    & -10.0$\pm$ 21.1 & 73.2   & 29.8$\pm$ 4.7   & 16.4 \\   
 Halo    &         &  3    & 27.2$\pm$ 63.6  & 110.1  &  21.5$\pm$14.2  &  24.6 \\   
H327+40  & 327/+40 &       &                   &      &                 &      \\   
 Disk    &         & 2     & 14.4$\pm$ 30.2  &  42.7  &  23.2$\pm$ 9.0  &  12.7 \\   
 ThDi    &         & 52   & -22.7$\pm$ 8.6   & 62.2   & 34.3$\pm$ 2.6   & 18.6\\   
 MWTD    &         & 24   &-20.7$\pm$ 9.3    &45.6    &33.7$\pm$ 2.8    & 13.7\\    
 Halo    &         & 15   & 70.4$\pm$ 30.2   & 117.0  & 6.3$\pm$ 9.0    & 35.0 \\   
P858     & 329/+32 &       &                   &      &                 &      \\   
 Disk    &         & 22    &-24.2$\pm$ 8.0   & 37.6   & 34.5$\pm$ 2.3   & 10.9 \\   
 ThDi    &         &  54   &-16.4$\pm$ 6.6   & 48.7   & 32.2$\pm$1.9    & 14.0 \\ 
 MWTD    &         &  18   &  6.0$\pm$18.4   & 77.9   & 25.8$\pm$ 5.3   & 22.6\\   
 Halo    &         &  9    & 78.2$\pm$ 32.6  &  97.7  &   4.8$\pm$ 9.4  &  28.3\\    
H330+20  & 330/+20 &      &                   &      &                 &      \\   
 Disk    &         &  3   & -8.9$\pm$ 43.4  &  75.1  &  29.9$\pm$11.5  &  19.9\\    
 ThDi    &         &  54  & -47.4$\pm$ 8.4  &  61.5  &  40.1$\pm$2.2   & 16.4 \\    
 MWTD    &         &  11  & -34.5$\pm$ 26.0 &   86.1 &  36.7$\pm$ 6.9  &  23.0\\   
 Halo    &         &   8  & -39.7$\pm$ 41.6 &  117.6 &   38.0$\pm$ 11.1 &  31.3 \\   
H333+37  & 333/+37 &      &                   &      &                 &      \\   
 ThDi    &         & 40   & -2.6$\pm$ 6.0   & 37.7   & 28.4$\pm$2.1    & 13.0 \\     
 MWTD    &         & 35   & -2.7$\pm$ 10.7  &  63.3  &  28.4$\pm$ 3.7  &  21.8 \\  
 Halo    &         & 20   & 36.5$\pm$ 33.2  & 148.5  &  15.0$\pm$11.4  &  50.9\\   
P741     & 339/+41 &      &                   &      &                 &      \\   
 ThDi    &         & 15   &-49.1$\pm$10.9   & 42.3   & 50.1$\pm$ 5.0   & 19.5 \\   
 MWTD    &         & 17   & -54.4$\pm$14.3  & 59.0   & 52.4$\pm$ 6.5   & 26.9 \\   
 Halo    &         &  2   &-126.9$\pm$ 29.0 & 41.0   & 86.1$\pm$ 13.4  &  19.0\\  
P802     & 339.5/+33&      &                   &      &                 &      \\   
 ThDi    &          & 65  & -32.1$\pm$ 6.8 &  55.0  & 41.1$\pm$ 2.9  &  23.3 \\   
 MWTD    &          & 31  & -14.4$\pm$11.1 & 61.6   & 33.6$\pm$ 4.7  &  26.1 \\    
 Halo    &          & 10  &  31.2$\pm$ 41.0 &  129.7& 14.2$\pm$ 17.4 &  55.1 \\   
\enddata 
\end{deluxetable}

\newpage

\begin{deluxetable}{lcccc}  
\tablenum{Fig10}
\tabletypesize{\scriptsize}
\tablewidth{0pt}
\tablecaption{Data for Figure 10} 
\tablehead{
\colhead{Distance Range}  &   \colhead{Mean Distance} &  \colhead{N$_{stars}$}  
& \colhead{{$<V_{LSR}>$}} & \colhead{Std. Error}\\ 
\colhead{kpc}   &  \colhead{kpc} &    &   \colhead{km s$^{-1}$}  &  \colhead{$\pm$ km s$^{-1}$}  
}
\startdata
$<$ 1.0     &   0.81   &    35   &  -10.7  &   7.8   \\
1.0 -- 1.5      &   1.30   &  132    &  -18.4  &   3.6  \\ 
1.5 -- 2.0    &   1.76   &  261    &  -15.7  &   3.0  \\
2.0 -- 2.5    &   2.23   &  279    &  -17.8  &   3.1  \\   
2.5 -- 3.0    &   2.73   &  213    &  -20.1  &   3.7  \\
3.0 -- 3.5    &   3.23   &  155    &  -13.1  &   4.4 \\
3.5 -- 4.0    &   3.73   &  94     &   -18.3  &  5.5 \\
4.0 -- 4.5    &   4.21   &  48     &   -6.6  &  7.8 \\
4.5 -- 5.0    &   4.71   &  28     &   -4.3   &  12.0 \\ 
\enddata
\end{deluxetable}

\begin{figure} 
\plotone{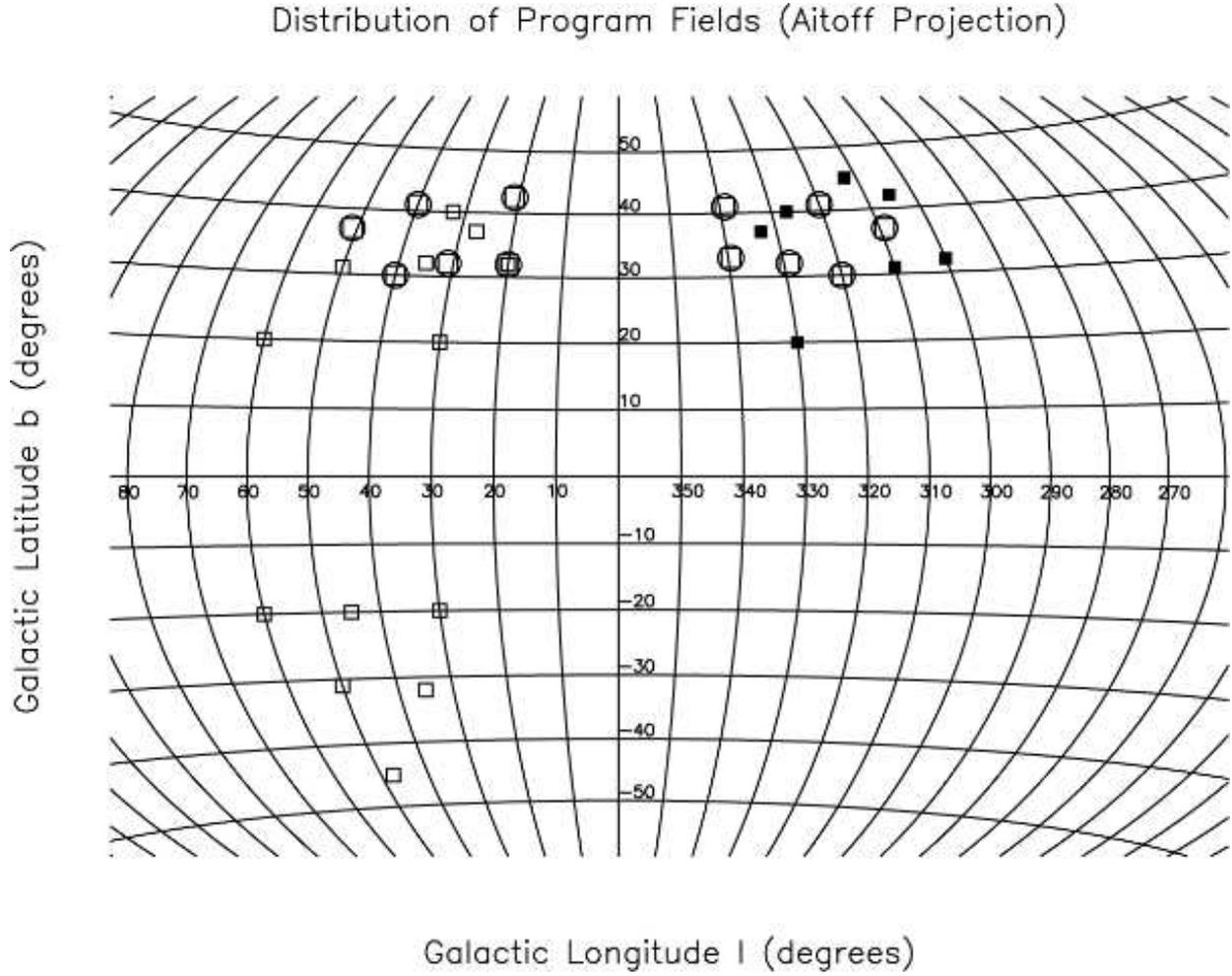}
\caption{Distribution of the spectroscopy fields. Open squares were observed 
with the MMT/Hectospec, filled squares with CTIO/Hydra, and the circles are from \citet{par04}}
\end{figure} 

\begin{figure}
\figurenum{2}
\epsscale{0.6}
\plotone{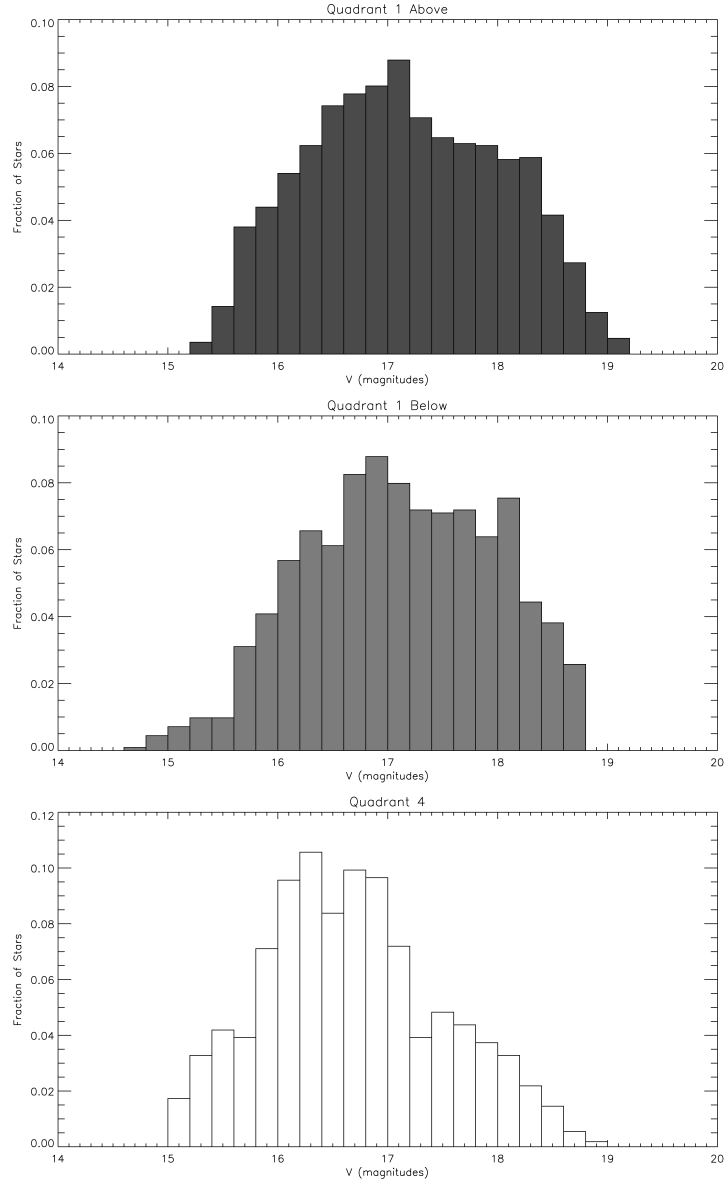}
\caption{Histograms of V magnitudes.} 
\end{figure}

\begin{figure}
\figurenum{3}
\epsscale{0.6}
\plotone{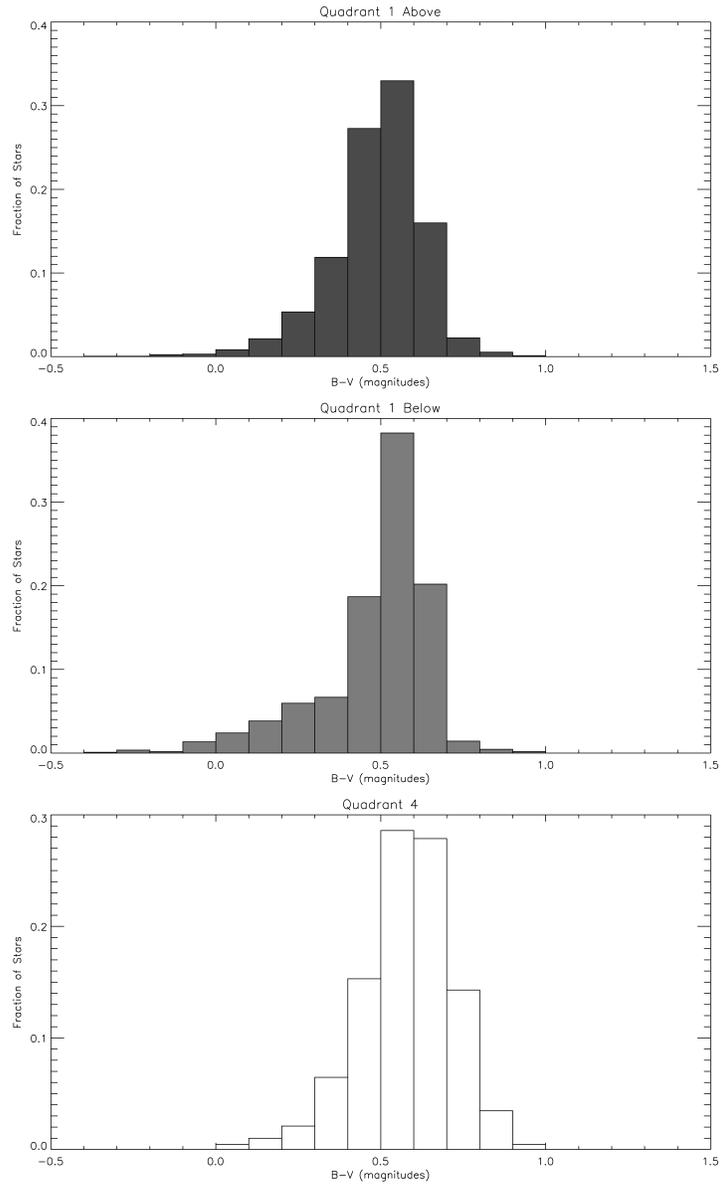}
\caption{Histograms of $B-V$ corrected for interstellar extinction. }
\end{figure}

\begin{figure}
\figurenum{4}
\epsscale{0.8}
\plotone{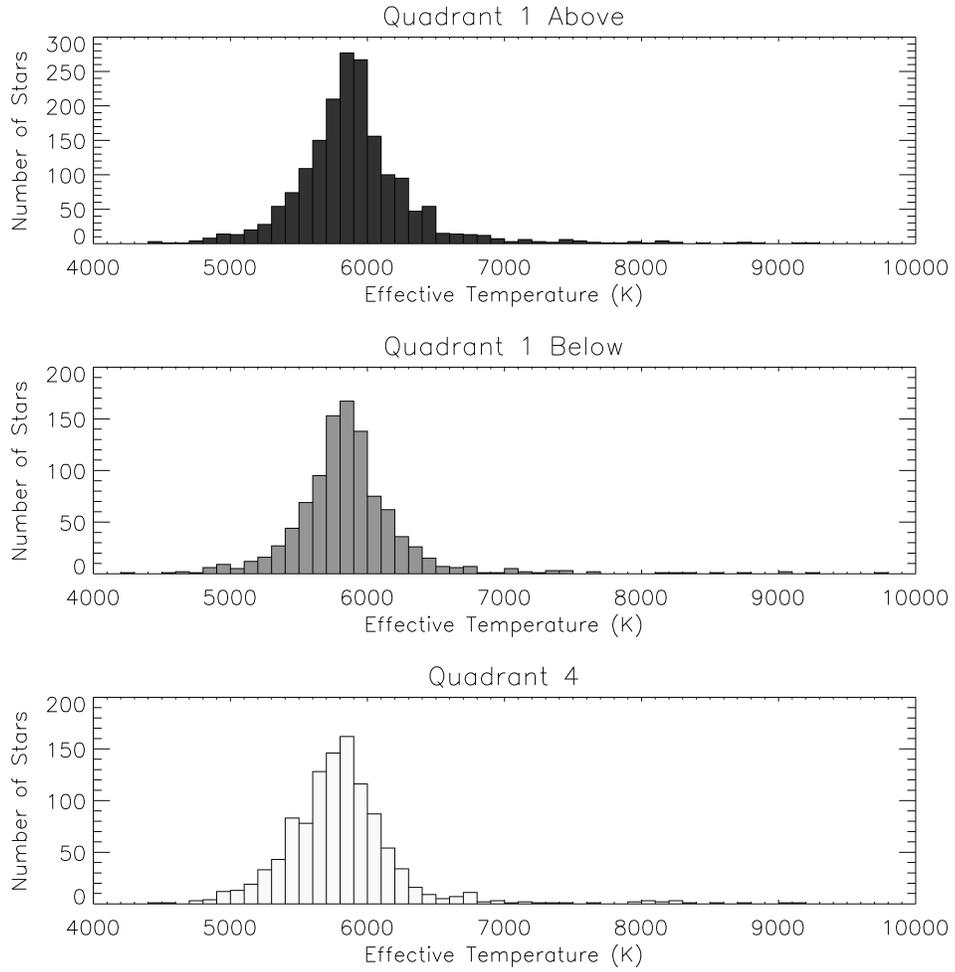}
\caption{Histograms of effective temperature from the n-SSPP pipeline.}
\end{figure}

\begin{figure}
\figurenum{5}
\epsscale{0.9}
\plotone{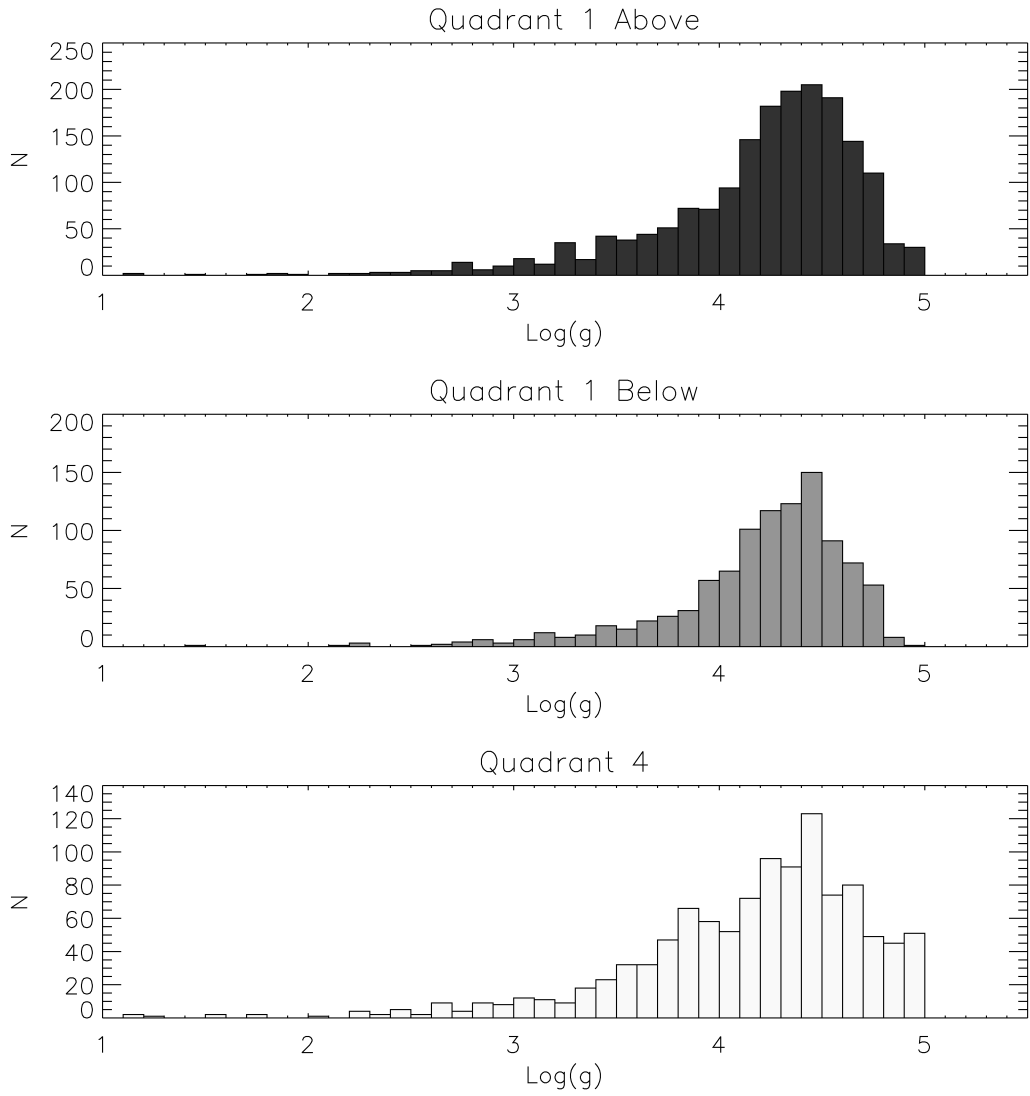}   
\caption{Histograms of log g (surface gravity) from the n-SSPP pipeline.}
\end{figure}

\begin{figure}
\figurenum{6}
\epsscale{0.6}
\plotone{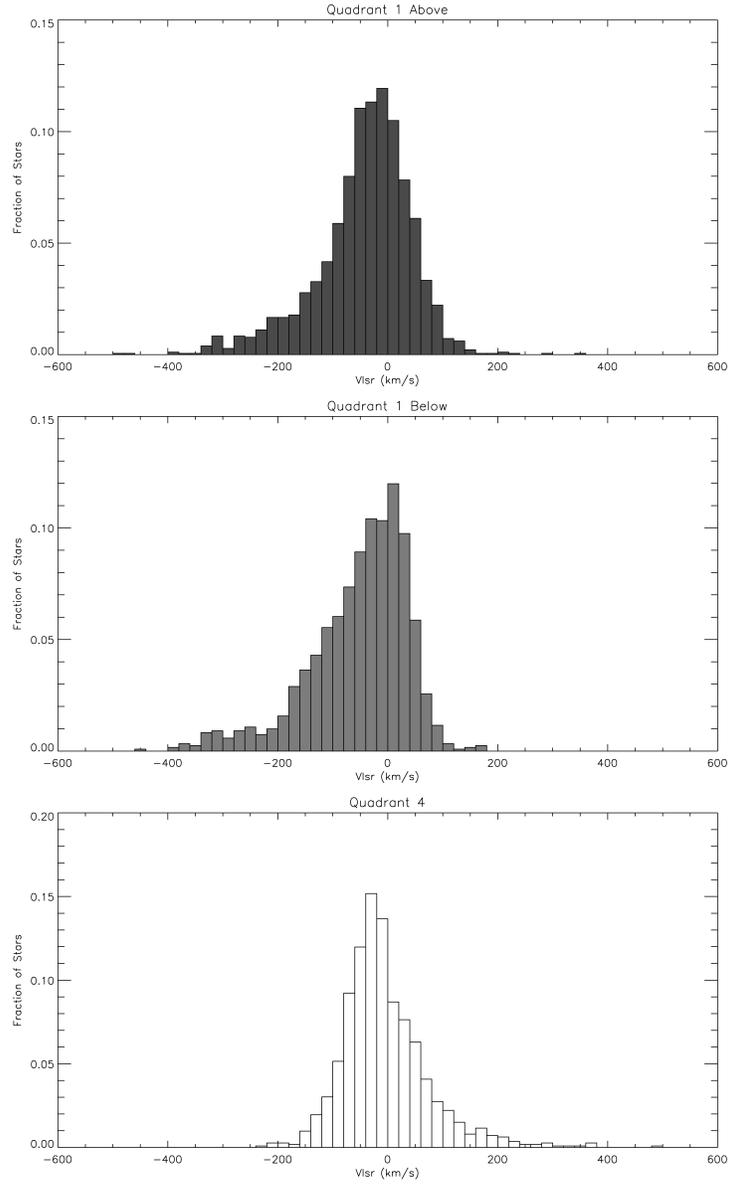}
\caption{The normalized velocity distribution for V$_{LSR}$.}
\end{figure}

\begin{figure}
\figurenum{7}
\epsscale{0.6}
\plotone{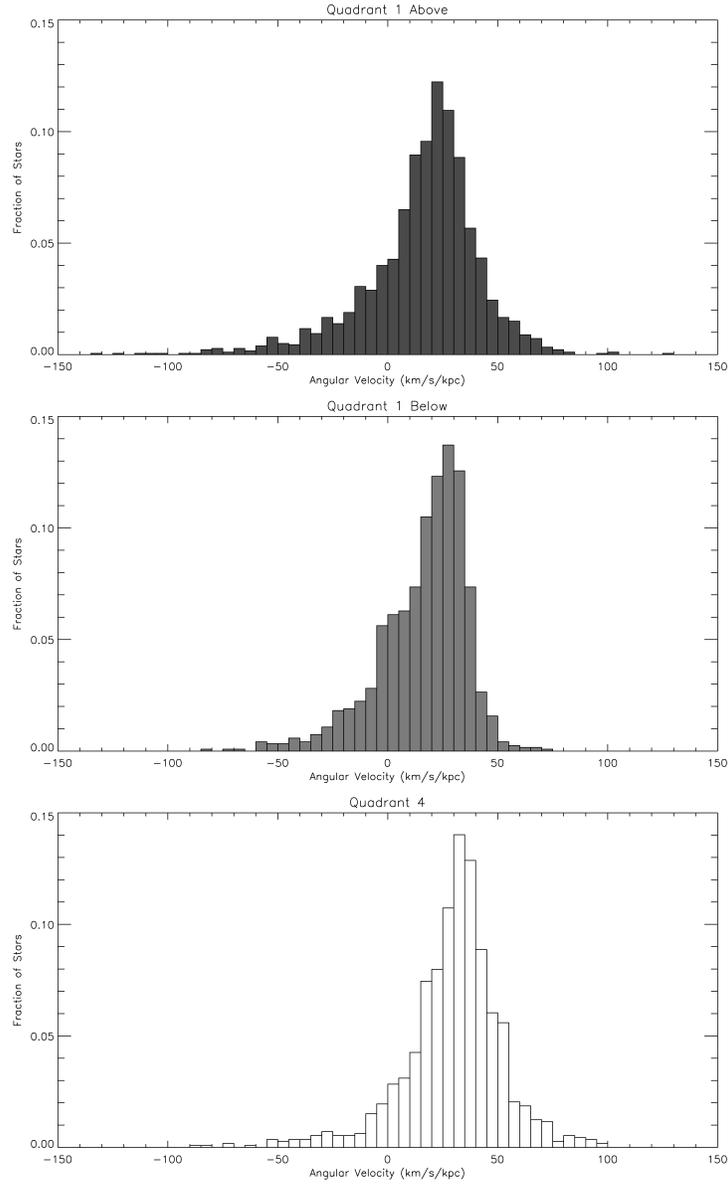}
\caption{The normalized distibution for the rotation rate $\omega$.}
\end{figure}

\begin{figure}
\figurenum{8}
\epsscale{1.0}
\plotone{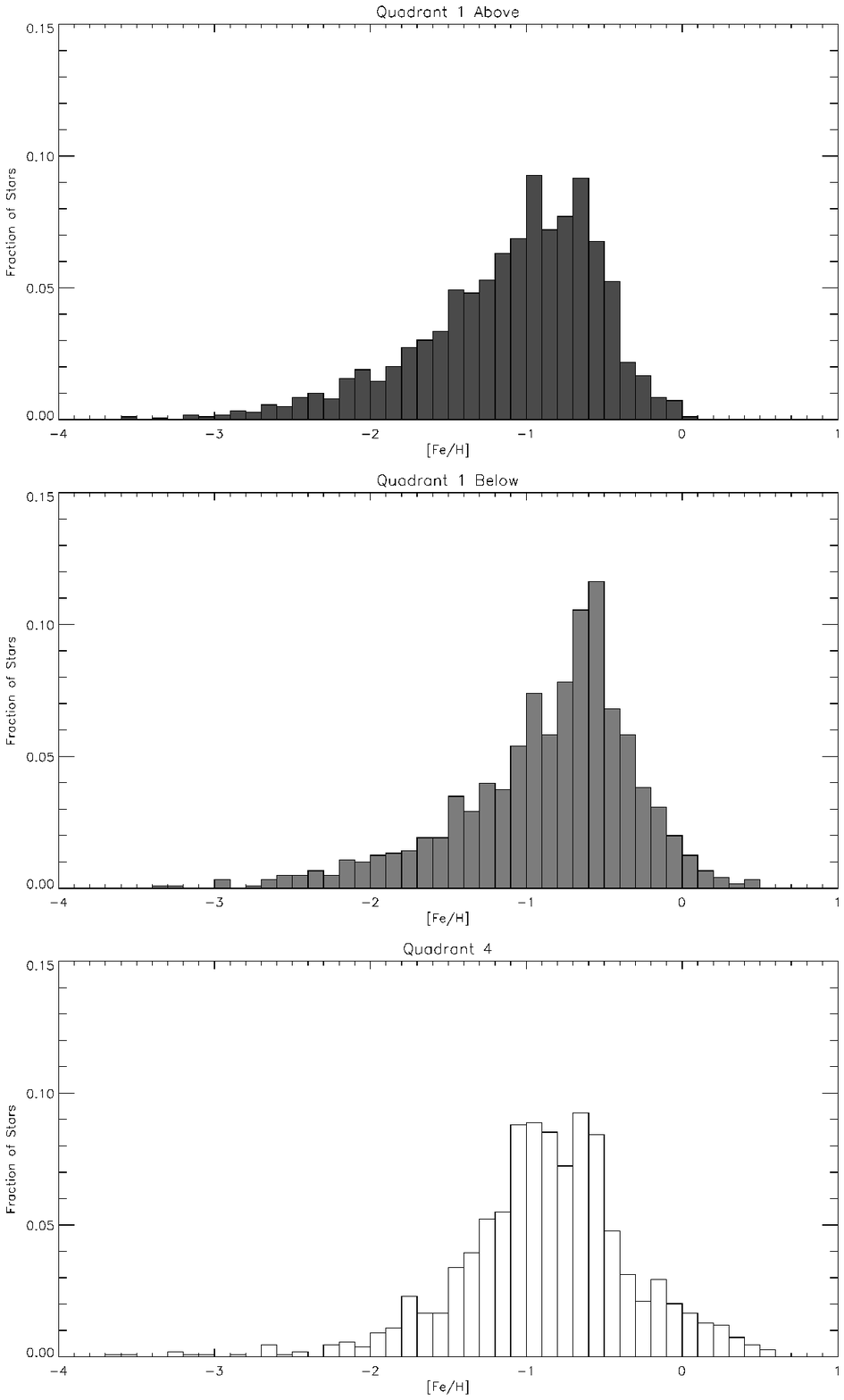}
\caption{The normalized metallicity distrinution functiond for [Fe/H].}
\end{figure}

\begin{figure}
\figurenum{9}
\epsscale{1.0}
\plotone{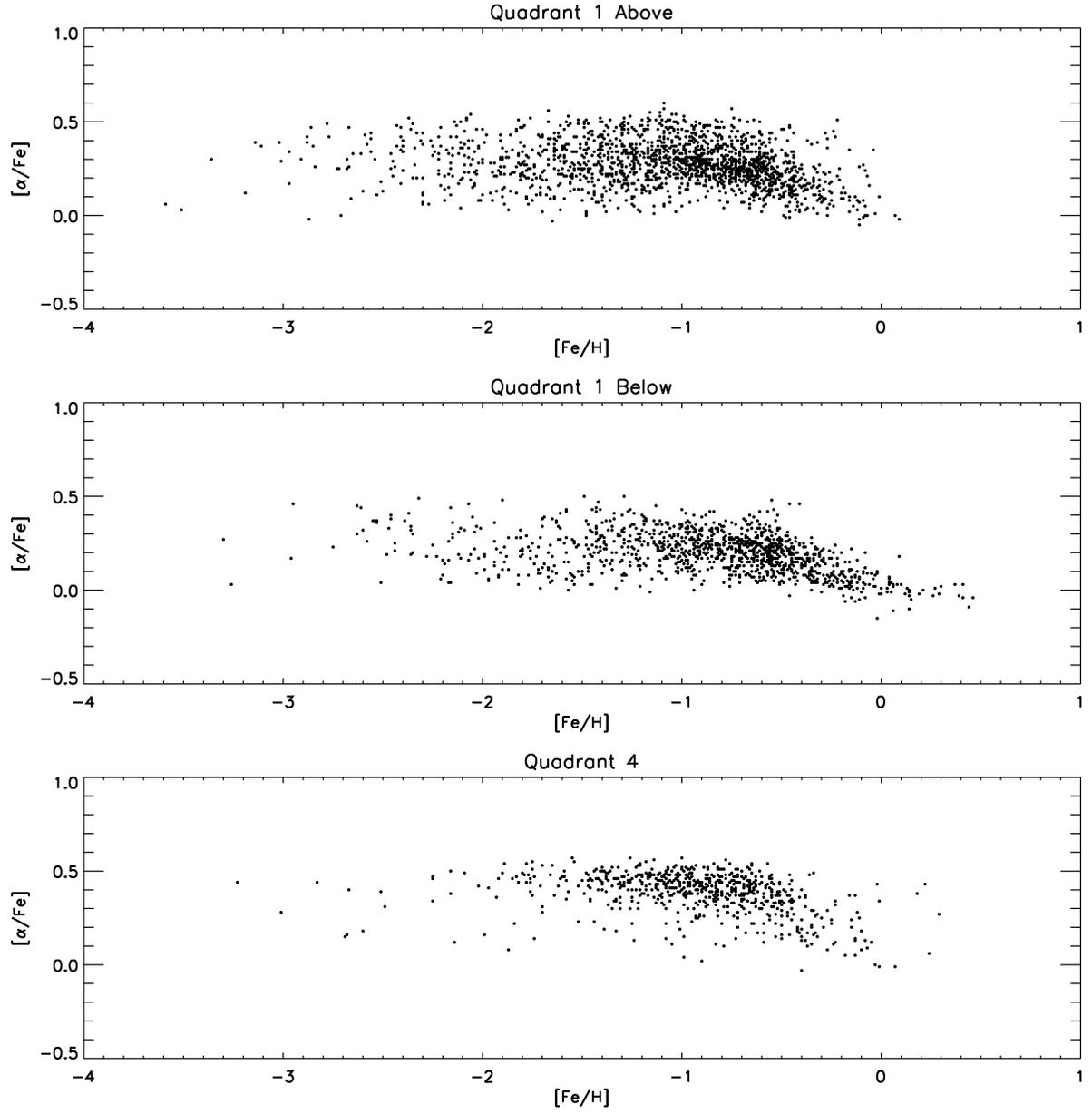}
\caption{The [$\alpha$/Fe] vs. [Fe/H] diagrams for the three regions.}  
\end{figure}



\begin{figure}
\figurenum{10}
\epsscale{1.0}
\plotone{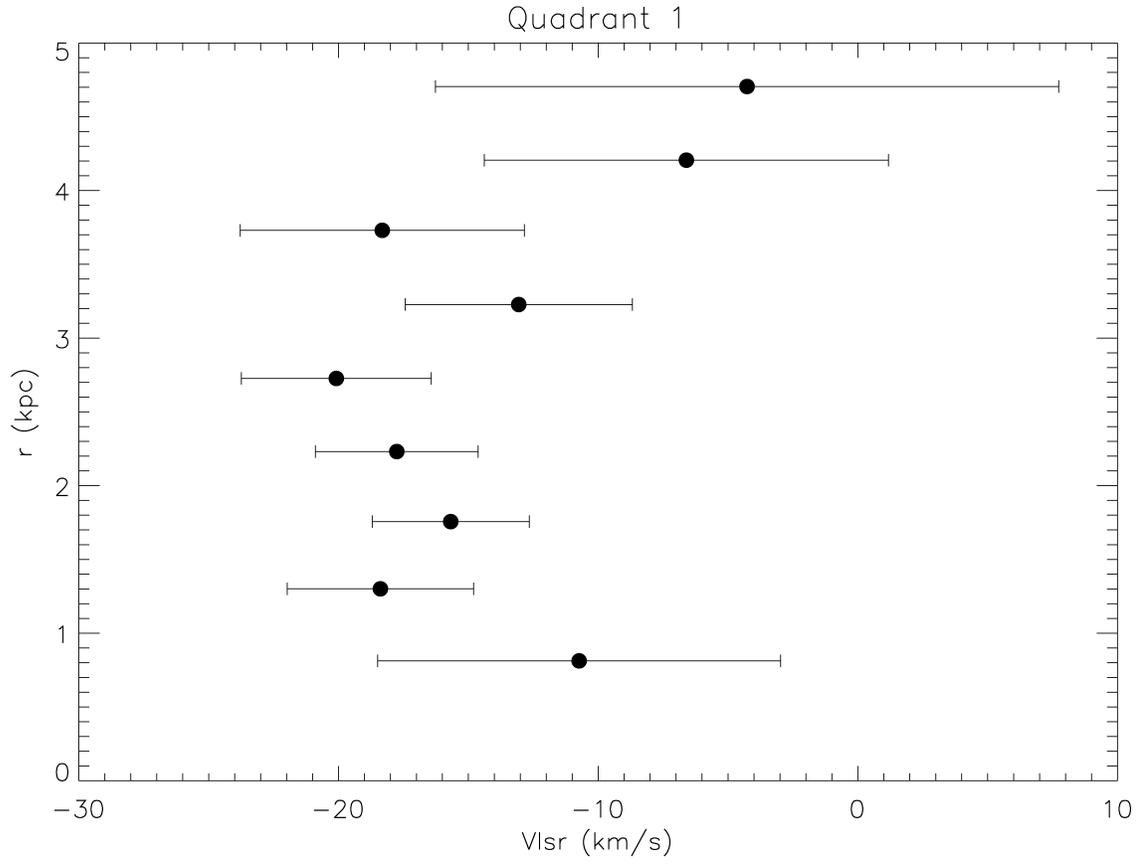} 
\caption{The mean V$_{LSR}$ velocity as a function of distance from the Sun for the Thick Disk stars
in in the first quadrant. Note the turnover or shift to less negative velocities at distances greater than 4 kpc.}
\end{figure}

\end{document}